\documentclass[a4paper]{aa}
\usepackage{ffbib}
\usepackage{times}
\usepackage{epsfig}

\newcommand{\tems}{\mbox{$EM$ vs.\ $T$ slope}}
\newcommand{\cq}{\mbox{$ \chi^2$}}
\newcommand{\zsun}{\mbox{$Z_\odot$}}

\begin{document}
\thesaurus{06(08.09.2 Algol; 08.12.1; 08.01.2; 08.03.05; 13.25.5)}

\title{Spectroscopic analysis of a super-hot giant flare observed on
  Algol by BeppoSAX on 30 August 1997}

\author{F. Favata\inst{1} \and J.\,H.\,M.\,M. Schmitt\inst{2} }

\institute{Astrophysics Division -- Space Science Department of ESA, ESTEC,
  Postbus 299, NL-2200 AG Noordwijk, The Netherlands
\and
Universit\"at Hamburg, Hamburger Sternwarte, Gojenbergsweg 112,
  D-21029 Hamburg, Germany 
}

\offprints{F. Favata} \mail{ffavata@astro.estec.esa.nl}

\date{Received 21 June 1999; accepted 4 August 1999}

\titlerunning{Spectroscopic analysis of the BeppoSAX Algol flare}
\authorrunning{F.~Favata \& J.\,H.\,M.\,M.~Schmitt}

\maketitle 
\begin{abstract}
  
  We present an X-ray observation of the eclipsing binary Algol,
  obtained with the BeppoSAX observatory. During the observation a
  huge flare was observed, exceptional both in duration as well as in
  peak plasma temperature and total energy release. The wide spectral
  response of the different BeppoSAX instruments, together with the
  long decay time scale of the flare, allowed us to perform a detailed
  time-resolved X-ray spectroscopic analysis of the flare.  We derive
  the physical parameters of the emitting region together with the
  plasma density applying different methods to the observed flare
  decay.  The X-ray emission from the flare is totally eclipsed during
  the secondary optical eclipse, so that the size of the emitting
  region is strongly constrained (as described in a companion paper)
  on purely geometrical arguments. The size of the flare thus derived
  is much smaller than the size derived from the analysis of the
  evolution of the spectral parameters using the quasi-static cooling
  formalism, showing that the time evolution of the flare is
  determined essentially from the temporal profile of the heating,
  with the intrinsic decay of the flaring loop having little
  relevance. The analysis of the decay with the technique recently
  developed for solar flares by \cite*{rbp+97} on the other hand is in
  much better agreement with the eclipse-derived constraints.
  
  The very high signal-to-noise of the individual spectra strongly
  constrains some of the derived physical parameters. In particular,
  very significant evidence for a three-fold increase in coronal
  abundance and for a large increase in absorbing column density
  during the initial phases of the flare evolution is present.


\end{abstract}

\section{Introduction}
\label{sec:intro}

Algol-type systems are short-period eclipsing binaries composed of an
early-type primary and a (near) main-sequence late-type secondary. The
short orbital period ensures that both stars are tidally-locked fast
rotators.  The secondaries in these systems show strong magnetic
activity with copious X-ray emission. While tidally-induced activity
is a common characteristic in RS~CVn-type systems, the early-type
primary in Algol-type systems, lacking a surface convective envelope,
is not expected to be able to sustain a dynamo. Therefore the coronal
structures should be concentrated on the secondary only, making for a
simpler geometry and avoiding the complication of cross-system loop
structures which may exist in the case of RS~CVn-type binaries.
Because of its proximity (28.46~pc on the basis of the Hipparcos
parallax, \cite{sp1200-1}) the eponymous system Algol is one of the
apparently strongest coronal X-ray sources in the sky.

Its brightness ($V \simeq 2.7$) soon made the periodic fading obvious
(it was first reported, in the western world, by Geminiano Montanari
in 1667), and their interpretation in terms of mutual eclipses in a
binary system was already proposed by John Goodricke in 1782.  The
system consists of a B8\,V primary and a K2\,IV secondary (plus a more
distant tertiary component, with a period of $\simeq 1.8$ yr and a
spectral type A or F).  \cite*{hbh+71} report values for the masses
and radii of the two components $R_{\rm A} = 3.0\,R_\odot$, $M_{\rm A}
= 3.7\,M_\odot$ and $R_{\rm B} = R_{\rm K} = 3.4\,R_\odot \simeq 2.4
\times 10^{11}$~cm, $M_{\rm B} = 0.8\,M_\odot$, while \cite*{ric93}
reports $R_{\rm A} = 2.90\,R_\odot$, $M_{\rm A} = 3.7\,M_\odot$ and
$R_{\rm B} = R_{\rm K} = 3.5\,R_\odot \simeq 2.5 \times 10^{11}$~cm,
$M_{\rm B} = 0.81\,M_\odot$; the orbital inclination is reported to be
$i = 81.4$~deg. We will adopt the \cite*{ric93} parameters in the
following. The orbital period is $\simeq 2.8673$~d.  The ephemeris we
have adopted in the present work is HJD 2\,445\,739.0030 +
2.8673285\,$E$ (\cite{kim89}; \cite{amf85}).  The separation is
14.14\,$R_\odot$, or $\simeq 4$ times the radius of the K star.

Algol was identified as an X-ray source already in the '70s with
observations from the SAS~3 satellite (\cite{sde+76}) and its soft
X-ray emission was confirmed with sounding rocket flights
(\cite{hft+77}). Its intense activity level has made it a target of
choice for most UV, EUV and X-ray observatories. The high level of
X-ray emission was initially interpreted in the framework of the
mass-transfer paradigm, given the evidence from optical data of mass
transfer taking place between the two components (see \cite{ric93} and
references therein).  However, spectroscopic observations soon showed
a hot, thermal X-ray spectrum, requiring the presence of magnetically
confined plasma, i.e.\ of a corona, expected to be located on or
around the K-type secondary.

The soft X-ray emission of Algol is characterized by the frequent
occurrence of major flares. Almost all sufficiently long observations
to date have yielded a significant flaring event, with EXOSAT
(\cite{wcp+86}; \cite{om89}), GINGA (\cite{sut+92}) and ROSAT
(\cite{os96}) all observing long-lasting flares (with effective decay
times ranging between 5 and 20~ks), which have been extensively
discussed in the literature.  The above flares have all been analyzed
in a similar way, allowing for a reasonably homogeneous comparison of
their characteristics to be made.  In particular, the observed decay
has in all cases been used to derive (following the formulation of
\cite{om89}) the length, and consequently, the average density of the
flaring plasma, under the assumption that the flaring loop evolves
throuh a series of ``quasi-static'' loop states. In all cases the
analysis has made use of the observed constancy of the normalized
ratio between the radiative and conductive cooling time $\mu$ (see
Sect.~\ref{sec:qs}) to ascertain that the flaring loop(s) are cooling
through a sequence of quasi-static states, and of the small value of
the heating function present as a parameter in the quasi-static
formalism to establish the lack of additional heating during the flare
decay phase.

The loop lengths\footnote{All along the present paper, the term ``loop
  length'' will be used to indicate the length from the footpoint to
  the apex of the loop, i.e.\ actually its ``semi-length''.} thus
derived range between $1$ and $6 \times 10^{11}$~cm (i.e.\ between 0.4
and 2.4 $R_K$ -- see Table~\ref{tab:comp}). These large loops
(comparable or larger in size than the stellar radius, unlike the
solar case, in which flaring structures are small compared to the
solar radius) are however shorter than the coronal pressure scale
height because of their high temperatures and the low surface gravity
of the K-type subgiant Algol secondary.  The corresponding plasma
densities, derived within the same framework, range between $5$ and
$26\times 10^{10}$~cm$^{-3}$.  The general picture for the flaring
regions observed on Algol is therefore one of large and rather tenuous
loops, a natural consequence of the very energetic and long-lasting
flares if no heating is indeed present during the decay phase. In the
solar case, in addition to the ``compact'' flares, in which the plasma
appears to be confined to a single loop whose geometry does not
significantly evolve during the event, a second class of flares is
usually recognized, i.e.\ the ``two-ribbon'' flares, in which a
disruptive event appears to open up an entire arcade of loops, which
subsequently closes back, leading to the formation of a growing system
of loops whose footpoints are anchored in H$\alpha$-bright
``ribbons''. Two-ribbon flares are generally characterized by a slower
rise and decay, and a larger energy release.  Compact flares have
often been considered to be due to ``impulsive'' heating events, while
the longer decay times of two-ribbon events have been considered as a
sign of sustained heating. However, also in the case of compact flares
sustained heating has been shown to be frequently present
(\cite{rbp+97}), so that the distinction may indeed be less clear than
often thought.

Long-lasting stellar flares have at times considered as analogs to
solar two-ribbon flares (due to their longer time scales, e.g.\ 
\cite{pts90}). However, the only available theoretical framework so
far available to model this type of event (\cite{kp84}) relies on a
large number of free parameters and assumptions (such as the
conversion efficiency of the magnetic field into X-rays and the
assumption of instantaneous dissipation of the reconnection energy).
As a consequence the physical parameters of the flaring regions
derived with this approach are, for spatially unresolved events such
as the stellar ones, rather strongly dependent on some specific
assumptions, so in practice most stellar flares have been modeled as
compact events. We will follow the same approach here, however keeping
in mind the possibility that the event may not be necessarily
described as a compact one.

We have performed a long ($\simeq 240$~ks elapsed time, covering a
full orbit of the system) observation of Algol with the BeppoSAX X-ray
observatory, aiming at studying both the spectral and the temporal
characteristics of its X-ray emission.  During the observation a very
large and long-lasting flare was observed. We present in this paper a
detailed analysis of the characteristics of this flare, deriving the
temporal evolution of the spectral parameters of the plasma
(temperature $T$, emission measure $EM$, coronal abundance $Z$,
absorbing column density $N({\rm H})$), and subsequently applying
different methods to the analysis of the flare decay in order to
derive the physical characteristics of the flaring region. For this
purpose we have applied both the quasi-static decay method of
\cite*{om89} and the method of \cite*{rbp+97}, which allows for the
possibility of (exponentially decaying) sustained heating during the
flare decay, and simultaneously deriving both the time scale of
heating and the size of the flaring loop.  In line with the previous
analyses of large flares on Algol, the analysis of the flare's decay
using the quasi-static approach results in a long and tenuous flaring
loop, although in this case the derived loop size and density are more
extreme given the exceptionally long duration and peak temperature of
the event.

One unique characteristic of the flare studied here is that its
emission underwent a total eclipse coincident with the secondary
optical eclipse. This allows (as discussed in detail in a companion
paper, \cite{sf99}) to put a firm upper limit to the size of the
flaring region, and thus to compare, for the first time in a context
other than the solar one, the length derived through the analysis of
the flare decay with the geometrical size of the emitting region.

This comparison shows that the loop sizes derived from the analysis of
the flare decay through the quasi-static method are significantly
larger than the geometrical size of the flaring region. Therefore the
actual flaring region must have had a much larger plasma density, and
sustained heating must have been present during all the decay phase to
explain the long observed decay time.  The method of \cite*{rbp+97}
produces a large range of allowed loop lengths, which at the lower end
overlap with the size derived for the flaring region from the eclipse
analysis. Also, this type of analysis points to the presence of
significant sustained heating during the decay phase.

The metal abundance of the flaring plasma is seen to vary
significantly during the course of the flare's evolution.  Abundance
variations during the evolution of the flare were already hinted at in
the analysis of the GINGA (\cite{sut+92}) and ROSAT (\cite{os96})
flares, and evidence for this type of effect has been reported for
flares on other stars.  The combination of high statistics, good
spectral resolution and wide spectral coverage of the present Algol
observation make it however possible for the first time to
quantitatively derive the evolution of the plasma abundance. Finally,
large variations of absorbing column density are also observed during
the early evolution of the flare, hinting at the possibility of a
coronal mass ejection taking place in association with the onset of
the flare.

Although flares of different types from several classes of coronal
sources have been discussed in detail in the literature in the past,
the large flaring event on Algol discussed in the present paper is
exceptional for several reasons:

\begin{itemize}
  
\item Its long duration (almost two days) associated with its high
  luminosity allows for high signal-to-noise spectra to be collected
  with a time resolution small compared with the time scale of flare
  evolution, and thus to analyze in detail the temporal evolution of
  the plasma parameters with small statistical errors, and on
  different time scales.
  
\item The complete, uninterrupted time coverage, from several thousand
  seconds before the onset of the flare until the end of its decay
  allows for its complete temporal evolution to be studied.
  
\item The occurrence of a \emph{total} eclipse of the flaring plasma
  by the primary star allows, for the first time, for a geometrical
  determination of the size of the flaring structure (see
  \cite{sf99}), which can then be compared with the loop lengths
  derived through an analysis of the decay of the spectroscopic
  parameters. This allows for a critical test of the assumptions
  underlying these techniques, which are the only ones available when
  confronted with stellar flares with no spatial resolution.
  
\item The high X-ray flux and spectral temperature associated with
  this event, together with the unprecedented energy coverage offered
  by the instrumental complement of the BeppoSAX observatory allow for
  the spectrum of the flaring plasma to be studied between 0.1 and
  100~keV, thus removing the uncertainties on the temperature of the
  flaring plasma during the hottest phases of large flares (where even
  the ASCA instruments can only provide lower limits to the
  temperature) which have characterized previous analyses of large
  flares. At the same time, the spectral shape can be critically
  determined, in particular looking for the presence of non-thermal
  spectral components.

\end{itemize}

\section{The BeppoSAX observation}
\label{sec:obs}

The BeppoSAX observatory (\cite{bbp+97}) features different
instruments, four of which were used in the analysis of our Algol
observation, i.e.\ the LECS (\cite{pmb+97}, which covers the energy
range 0.1--10~keV), the two MECS detectors (\cite{bcc+97}, MECS-2 and
MECS-3, covering the range 1.6--10~keV) and the PDS detector
(\cite{fcd+97}, covering between 15 and $300$~keV -- only data in the
15--100~keV band were used in the present paper).
 
The BeppoSAX observation of Algol covered a complete binary orbit
(i.e.\ $\simeq 240$\,ks elapsed time). It started on Aug.\ 30, 1997 at
03:04 UT (shortly before the primary optical eclipse) and lasted until
Sep.\ 1, 1997 at 20:32 UT. Approximately 20\,ks after the beginning of
the observation, a very strong flare began, whose evolution dominates
the rest of the observation.

A detailed analysis of the total eclipse of the flare as seen in the
MECS detectors is presented by \cite*{sf99}, who derive the
corresponding geometrical constraints on the size and shape of the
flaring region. The present paper will concentrate on the spectral
analysis of the X-ray emission and on the analysis of the
characteristics of the flaring region from the flare decay, using the
complete spectral range covered by the BeppoSAX detectors.

\begin{figure*}[htbp]
  \begin{center}
    \leavevmode \epsfig{file=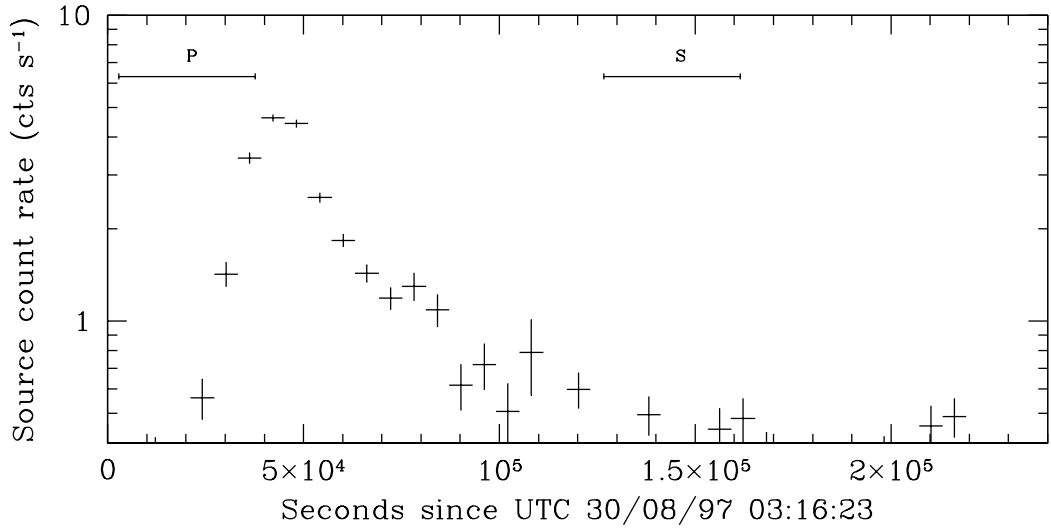, width=15.0cm, bbllx=20pt,
      bblly=400pt, bburx=600pt, bbury=700pt, clip=}
    \caption{The light-curve of the Algol observation of BeppoSAX in the
      15--100~keV band, as measured in the PDS detector, binned in
      6000\,s intervals.  The location and duration of the primary and
      secondary optical eclipse, using the ephemeris of Kim (1989), is
      also plotted.}
    \label{fig:lcpds}
  \end{center}
\end{figure*}

\subsection{Data reduction}

Telemetry files, containing both information on individual detected
X-ray photons and house-keeping data were obtained from the
observation tapes for each instrument, and data for each instrument
were individually processed with the {\sc saxdas} software (available
from the BeppoSAX Scientific Data Center -- hereafter SDC, reachable
at {\tt http://www.sdc.asi.it}), with the default settings, producing
FITS-format linearized photon event files for each instrument.

For the three imaging instruments (LECS, MECS-2, MECS-3) standard
extraction regions were used, i.e.\ 8.2 and 4.0~arcmin diameter
circles centered on the source, for the LECS and MECS data,
respectively. The background was extracted from regions of the same
size and location as the source extraction region from the standard
background files supplied by the SDC. Spectra and light-curves both
for the source and the background were extracted using the {\sc
  xselect} software.  PDS spectra and light-curves were extracted
using the {\sc saxdas}-supplied packages, which directly produce
background-subtracted spectra and light-curves.

\subsection{Light curves}

\begin{figure*}[htbp]
  \begin{center}
    \leavevmode \epsfig{file=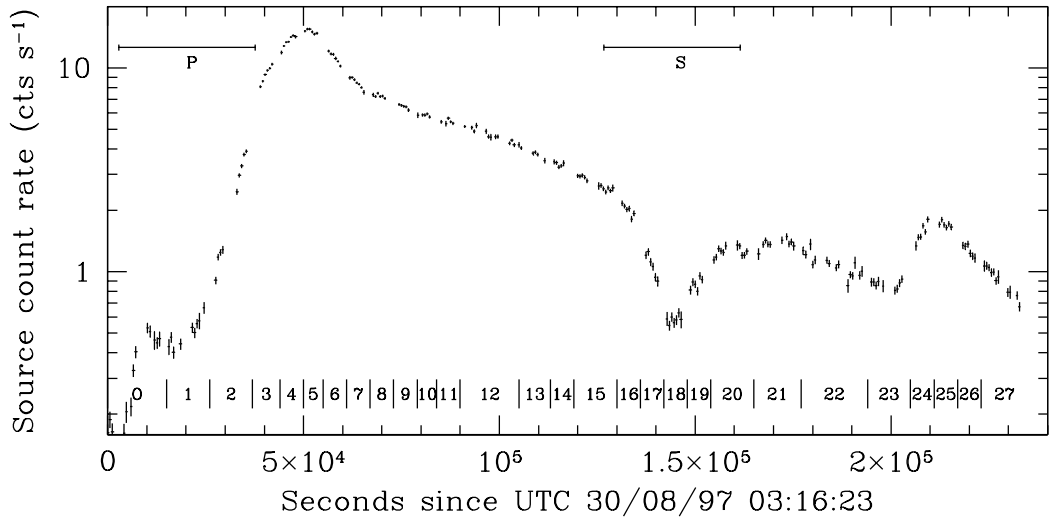, width=15.0cm, bbllx=20pt,
      bblly=400pt, bburx=600pt, bbury=700pt, clip=}
    \caption{The light-curve of the Algol observation of BeppoSAX in the
      1.6--10~keV band, as measured in the MECS-3 detector, binned in
      600\,s intervals. The vertical lines indicates the boundaries of
      the time segments in which the observation has been broken for
      the purpose of performing time-resolved spectroscopy. The
      location and duration of the primary and secondary optical
      eclipse, using the ephemeris of Kim (1989), is also plotted.}
    \label{fig:lc}
  \end{center}
\end{figure*}

\begin{figure*}[htbp]
  \begin{center}
    \leavevmode \epsfig{file=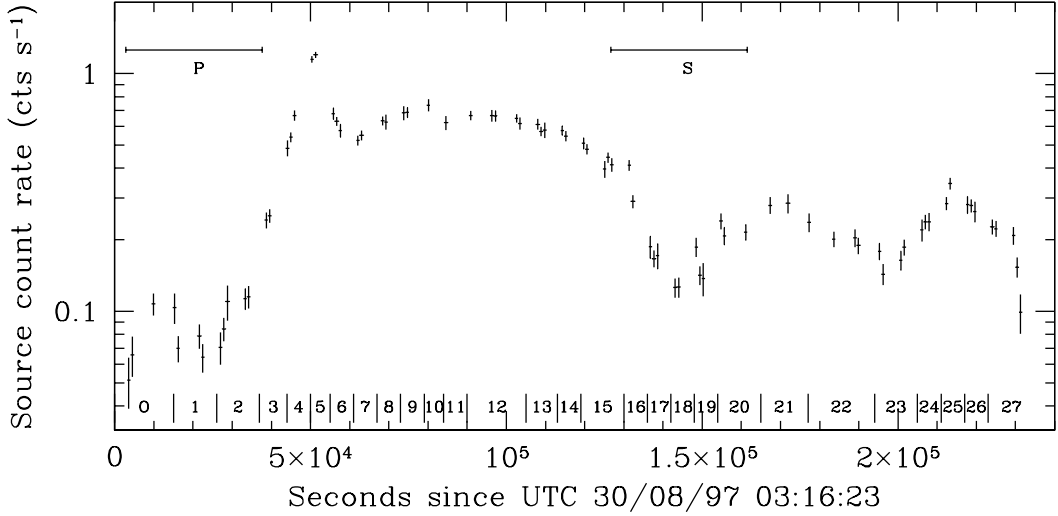, width=15.0cm, bbllx=20pt,
      bblly=400pt, bburx=600pt, bbury=700pt, clip=}
    \caption{The light-curve of the Algol observation of BeppoSAX in the
      0.1--0.5~keV band, as measured in the LECS detector, binned in
      900~s intervals. The vertical lines indicates the boundaries of
      the time segments in which the observation has been broken for
      the purpose of performing time-resolved spectroscopy. The
      location and duration of the primary and secondary optical
      eclipse, using the ephemeris of Kim (1989), is also plotted.}
    \label{fig:lcle}
  \end{center}
\end{figure*}

The background-subtracted light-curve in the 15--100~keV band,
extracted from the PDS data, is shown in Fig.~\ref{fig:lcpds}, binned
in 6000~s intervals, while the background-subtracted light-curve in
the 1.6--10.0~keV band, extracted from the MECS-3 detector is shown,
binned in 600~s intervals, in Fig.~\ref{fig:lc}. The light-curve for a
softer band (0.1--0.5~keV) derived from the LECS data is shown in
Fig.~\ref{fig:lcle}, binned in 900~s intervals.  The LECS is operated
during Earth night only, resulting in a lower observing efficiency and
in the larger data gaps seen in Fig.~\ref{fig:lcle} with respect to
Fig.~\ref{fig:lc}.

Inspection of Fig.~\ref{fig:lc} shows that the flare (as seen in the
1.6--10~keV band) has a rather slow rise (with $\simeq 30$~ks between
the flare start and the peak. The decay is for the first $\simeq
15$~ks relatively rapid, on time scales comparable with the rise, but
then slows down becoming very nearly exponential.  The eclipse due to
secondary is well visible, between $\simeq 130$ and $\simeq 160$~ks
from the beginning of the observation. The exponential decay is
interrupted, at $\simeq 200$~ks, by the onset of yet another flare.

The hard (15--100~keV) X-ray light curve shows a slow rise, similar to
the one seen in the 1.6--10~keV band, but a faster decay. The
hard X-ray count rate returns to its pre-flare value at $\simeq
130$~ks from the beginning of the observation, so that the eclipse of
the flaring plasma is not visible in this band. Also, there is no
evidence for hard pre-flare emission, which could in principle have
been due to non-thermal emission associated with fast particles.

The soft-band (0.1--0.5~keV) light curve differs significantly from
the 1.6--10~keV band light curve, as the rise is slower, and the decay
``bounces back'', after $\simeq 10$~ks, for $\simeq 20$~ks.  These
differences are discussed in more detail in Sect.~\ref{sec:abs}.

\begin{table*}[htbp]
  \begin{center}
    \caption{The spectral parameters derived for the quiescent
      emission of Algol from the joint analysis of the LECS and MECS
      spectra, determined on the spectrum accumulated during the time
      intervals 0, 1 and 18. The fit had a reduced \cq\ of 0.93, with
      659 degrees of freedom. For each best-fit parameter the
      confidence interval is given as computed using the criterion
      $\Delta \chi^2 = 3.50$, corresponding to the 68\% confidence
      level in the case of three interesting parameters.} \leavevmode
    \begin{tabular}{rr|rr|rr|rr|rr|rr}
      \multicolumn{2}{c|}{$T_1$} & \multicolumn{2}{c|}{$T_2$} &
 \multicolumn{2}{c|}{$EM_1$} & \multicolumn{2}{c|}{$EM_2$} & 
 \multicolumn{2}{c|}{$Z$} & \multicolumn{2}{c}{$N({\rm H})$} \\ 
 \multicolumn{2}{c|}{keV}  & \multicolumn{2}{c|}{keV}  &
 \multicolumn{2}{c|}{$10^{52}$ cm$^{-3}$} &
 \multicolumn{2}{c|}{$10^{52}$ cm$^{-3}$} &  
 \multicolumn{2}{c|}{$Z_\odot$} & \multicolumn{2}{c}{$10^{20}$
   cm$^{-2}$} \\\hline  
 0.97 & [0.89 -- 1.1] & 3.2 & [2.5 -- 3.8] & 4.1 & [3.2 -- 5.0] & 25. &
 [19. -- 31.] & 0.37 & [0.24 -- 0.50] & 0.91 & [0.54 -- 1.3]\\
\end{tabular}
\label{tab:qui}
\end{center}
\end{table*}

\subsection{Spectra}

The high-count rate of the Algol observation in the LECS and MECS
detectors allows for detailed time-resolved spectroscopy. The
observation has therefore been split in 28 separate time intervals,
with boundaries coinciding with observational gaps due to the Earth
blocking the source, and with each segment covering one or more
spacecraft orbits. The extent of each time segment is shown, together
with a number used in the following to refer to them, in
Figs.~\ref{fig:lc} and~\ref{fig:lcle}, together with the light-curve
of the observation as measured in the MECS-3 and LECS detectors
respectively. The time of optical eclipses is also indicated.

Individual spectra have been extracted, in the LECS, MECS and PDS
detectors, for each of the time intervals indicated in
Fig.~\ref{fig:lc}. Standard response matrices have been used to
analyze the MECS spectra (again as available from the SDC), while the
LECS response matrix was computed using the {\sc lemat} program. The
standard PDS pipeline already produces background-subtracted spectra
and thus no further manipulation was necessary. The standard
SDC-provided response matrix has been used to fit the PDS spectra.

All the spectral analysis described here has been performed using the
{\sc xspec} version 10.00 software. Each of the individual LECS and
MECS spectra accumulated during the intervals marked in
Fig.~\ref{fig:lc} has been rebinned, prior to fitting, to have at
least 20 counts per (variable-size) bin, and the statistical weight of
each bin for the purpose of determining the \cq\ has been determined
using the statistic of \cite*{geh86}, more appropriate than the
Gaussian approximation for small number of events.

Due to a know discrepancy between the normalization of the response
matrices for the LECS and MECS detectors, when LECS and MECS data are
fit together it is necessary to add a relative normalization factor to
the fit, with experience showing that the MECS normalization is about
10 to 20\% higher than the LECS one. It is not possible however to
determine a priori the exact value of the normalization, as this is a
function of the source position in the field of view as well as of the
source spectrum. We have therefore first performed a set of
two-temperature fits on the time-resolved spectra, leaving the
relative normalization of the LECS and MECS detectors as an additional
free parameters. The average value of the MECS to LECS normalization
thus determined is 1.15, with a range from 1.03 up to 1.24. The
behavior of the fit parameters and of the quality of the fits was then
determined by performing the same fits with the relative normalization
fixed to the average value of 1.15. A comparison of the probability
levels of the two sets of fits shows that leaving the relative
normalizations free to vary does not improve the fits, and all the
fits discussed in the following were thus performed with the relative
normalization fixed at 1.15.

All of the spectra discussed here have been fitted with a combination
of absorbed thermal emission models. The plasma emission code used has
in all cases been the {\sc mekal} model as present in {\sc xspec},
which implements the optically-thin, collisional ionization
equilibrium emissivity model described in \cite*{mkl95}. The presence
of absorbing material has been accounted for using a {\sc wabs}
component in {\sc xspec}, which implements the \cite*{mm83} model of
X-ray absorption from interstellar material. The metal abundance of
the emitting plasma is considered as a free parameter. Abundance
values are in the following determined as relative to the ``solar''
abundance, as determined by \cite*{ag89}. While the global metal
abundance was left free to vary, abundance ratios were kept fixed in
the fitting process.

\section{The quiescent spectrum}

Most of the BeppoSAX observation of Algol is occupied by the large
flare, and the quiescent emission is visible only in a small time
interval, i.e.\ before the flare itself (during the intervals marked 0
and 1) and during the total phase of the eclipse of the flaring plasma
(interval 18). These three spectra have been individually fit with a
two-temperature absorbed spectral model with freely varying global
metal abundance. Their best-fit parameters are the same within the
$1\,\sigma$ confidence range, and they have therefore been summed
together in a single spectrum representative of the average
``quiescent'' emission, from which average spectral parameters have
been determined, as listed in Table~\ref{tab:qui}.

Although significantly colder than the flaring emission, the plasma
responsible for the ``quiescent'' emission of Algol still has a high
temperature; the 3.2~keV (44~MK) observed here are somewhat higher
than the $\simeq 2.5$~keV (29~MK) reported on the basis of ASCA
observations by \cite*{anw94}.

The spectral parameters of the quiescent emission from the ROSAT PSPC
observation of Algol (which has a good out-of-flare phase coverage)
show the presence of orbital modulation (\cite{os96}), and thus a
proper subtraction of the quiescent emission from the flare spectra
would require this effect to be taken into account. Given the very
scant phase coverage of the quiescent spectra for the BeppoSAX
observation, however, we cannot determine the phase dependence
of the spectral parameters of the quiescent emission (if present).
The quiescent spectrum has therefore assumed to be constant,
with parameters given by the average of the best-fit parameters of the
three quiescent spectra. This average quiescent spectrum has been
added as a constant (``frozen'') model component in our subsequent analysis
of the flare spectra.

\begin{table*}[htbp]
  \begin{center}
    \caption{The set of best-fit parameters to the flaring emission,
      determined through the fitting process described in the text.
      The first column gives the number of the time interval as plotted
      in Fig.~\ref{fig:lc}, the second column the time from the
      beginning of the observation. The last column provides the number of
      degrees of freedom for each fit. For the best-fit temperature,
      metal abundance and absorbing column density the confidence
      interval is given, computed using the criterion $\Delta
      \chi^2 = 3.50$, corresponding to the 68\% confidence level in
      the case of three interesting parameters. The emission measure
      is given in units of $10^{54}$~cm$^{-3}$.  }
    \leavevmode
    \begin{tabular}{r|r|rr|r|rr|rr|r|r}
      Int. &  $t$ &    \multicolumn{2}{c|}{$T$} & $EM_{54}$ &
      \multicolumn{2}{c|}{$Z$} & \multicolumn{2}{c|}{$N({\rm H})$} &
      \cq & DOF \\  
      & ks & \multicolumn{2}{c|}{keV}  & cm$^{-3}$ &
      \multicolumn{2}{c|}{$Z_\odot$} &
      \multicolumn{2}{c|}{$10^{20}$~cm$^{-2}$} & & 
      \\\hline 
      2  &    31.5  &   12.37  & [10.9 -- 14.1] & 1.83  &  0.60  &
      [0.46 -- 0.76] &       26.8  & [22. -- 32.] & 0.927 & 615 \\ 
      3  &    40.5  &   10.96  & [10.4 -- 11.6] & 7.85  &   1.0  &
      [0.94 -- 1.1]  &      11.0  & [10. -- 13.] & 0.911 & 615 \\  
      4  &    47.0  &   10.14  & [9.8 -- 10.5]  & 11.4  &   1.1  &
      [1.0 -- 1.2]   &     4.03  & [3.8 -- 4.6] &   1.060 & 615 \\  
 5  &    52.5  &   8.56  &  [8.3 -- 8.8]   & 13.3  &  0.99  & [0.94 -- 1.0]  &
     1.60  &  [1.5 -- 1.8] & 1.293 & 615 \\ 
 6  &    58.0  &   6.76  &  [6.6 -- 7.0]   & 10.9  &  0.78  & [0.73 -- 0.82] &
     3.32  &  [3.2 -- 3.8] & 1.399 & 615 \\ 
 7  &    64.0  &   6.98  &  [6.8 -- 7.2]   & 8.13  &  0.66  & [0.61 -- 0.71] &
     3.12  &  [2.9 -- 3.6] & 1.038 & 615 \\ 
 8  &    70.0  &   7.19  &  [6.9 -- 7.5]   & 6.83  &  0.59  & [0.53 -- 0.65] &
     1.72  &  [1.5 -- 2.0] & 0.998 & 615 \\ 
 9  &    76.0  &   6.72  &  [6.4 -- 7.1]   & 6.33  &  0.46  & [0.40 -- 0.52] &
     1.23  &  [0.98 -- 1.6] & 0.953 & 615 \\ 
10  &    81.5  &   5.97  &  [5.7 -- 6.3]   & 5.95  &  0.45  & [0.38 -- 0.51] & 
     1.02  &  [0.72 -- 1.4] & 0.887 & 577 \\ 
11  &    87.0  &   5.73  &  [5.5 -- 6.0]   & 5.65  &  0.36  & [0.30 -- 0.43] &
     1.03  &  [0.68 -- 1.4] & 0.87 & 615 \\ 
12  &    97.5  &   5.37  &  [5.2 -- 5.5]   & 4.96  &  0.39  & [0.34 -- 0.42] &
     0.852  &  [0.76 -- 1.0] &  1.010 & 615 \\ 
13  &   109.0  &   4.77  &  [4.6 -- 5.0]   & 4.20  &  0.36  & [0.29 -- 0.43] &
     0.794  &  [0.61 -- 0.99] & 0.884 & 615 \\ 
14  &   116.0  &   4.36  &  [4.2 -- 4.6]   & 3.79  &  0.34  & [0.26 -- 0.42] &
     0.763  & [0.55 -- 0.97] &  0.863 & 615 \\ 
15  &   124.5  &   4.27  &  [4.1 -- 4.4]   & 2.91  &  0.40  & [0.34 -- 0.46] &
     0.620  & [0.55 -- 0.97] &  1.013 & 615 \\ 
21  &   171.0  &   3.35  &  [3.2 -- 3.6]   & 1.43  &  0.26  & [0.15 -- 0.37] &
     0.809  & [0.50 -- 1.8] & 0.771 & 550 \\
22  &   185.5  &   2.83  &  [2.7 -- 3.0]   & 1.00  &  0.50  & [0.35 -- 0.65] &
     1.27  & [0.85 -- 1.7] & 0.826 & 567 \\
    \end{tabular}
    \label{tab:pars}
  \end{center}
\end{table*}

\subsection{The H column density}

\cite*{wvv90} report an upper limit, to the H column density toward
Algol, of $2.5 \times 10^{18}$~cm$^{-2}$ (based on an upper limit to
the equivalent width of interstellar Na\,{\sc i} lines);
\cite*{sls+95} assume (based on these data) a value of $2 \times
10^{18}$~cm$^{-2}$ in their analysis of the Algol EUVE spectrum.
However, \cite*{os96} find, from the analysis of the ROSAT PSPC
spectra, a value of H column density higher by approximately one dex
(1 to $2 \times 10^{19}$~cm$^{-2}$). Higher values are found (again
from PSPC data) by \cite*{sdw95b}, who report $3.9\pm 1.0 \times
10^{19}$~cm$^{-2}$, and by \cite*{ott94} who report a range of values
(varying with the orbital phase) ranging between 3 and $8 \times
10^{19}$~cm$^{-2}$.  All X-ray derived values of the absorbing column
density are thus significantly higher than the values derived from the
EUVE data, and the BeppoSAX quiescent spectrum is no exception, with a
68\% confidence range of 5.4--$13 \times 10^{19}$~cm$^{-2}$,
comparable with the PSPC derived range of values of \cite*{ott94}.

Similar discrepancies between the X-ray and EUV-derived H column
densities are also observed for other active stars. For example, the H
column density toward the RS~CVn system AR~Lac has been estimated
(using the ratio of the 335 and 361~\AA\ Fe\,{\sc xvi} lines in the
EUVE spectrum) by \cite*{gj98} at $\simeq 2 \times 10^{18}$~cm$^{-2}$,
while \cite*{os94} derive, from the PSPC spectrum, a value of $\simeq
3 \times 10^{19}$~cm$^{-2}$. The value derived from the BeppoSAX
spectrum is $\simeq 6 \times 10^{19}$~cm$^{-2}$ (\cite{rpl+99}).

We have tried to fit the quiescent X-ray spectrum with a fixed, low H
column density. In practice, any value of $N({\rm H})$ lower than few
parts in $10^{19}$~cm$^{-2}$ results in fits with significantly higher
plasma metal abundance than the $1/3~Z_\odot$ obtained by leaving
$N({\rm H})$ as a free parameter: to balance the higher soft
continuum, the fit process increases the metal abundance, so that the
emission from the Fe\,L complex also increases (see \cite{fmp+97}). As
a result, however, the emission from the Fe\,K complex seen in the
MECS detector becomes significantly over-predicted.

To assess the significance of the above discrepancies we have
simulated the ability of LECS spectra to determined low values of the
H column density and the impact of possible calibration errors on the
best-fit column density. Even assuming a perfect calibration, with the
$S/N$ of the present spectra column densities of $\la
10^{19}$~cm$^{-2}$ cannot be distinguished from zero column densities
(although above $10^{19}$~cm$^{-2}$ their effect becomes visible, see
above).  The systematics in the LECS calibration are estimated to be
at the level of $\la 5$\%, although they could be higher (up to
$\simeq 10$\%) in the low-energy (0.1--0.2~keV) range, where the very
small absolute area and its steep slope as a function of energy make
the calibration very difficult (A.~Parmar, private communication).
The best-fit column density is very sensitive to possible calibration
errors at low-energies: our simulations show that a $\simeq 10$\%
calibration error in the spectral region below 0.2~keV could lead to a
systematic bias in the best-fit column density for stellar spectra of
few $10^{19}$~cm$^{-2}$, comparable with the best-fit column density
for the quiescent spectrum. Thus, the absolute value LECS-derived
best-fit column densities of this order should be regarded with
caution as they critically depend on the calibration being known to
better than the currently estimated uncertainty.  Relative changes in
the absorbing column density (as seen during the flare decay here)
would not be affected by such errors, nor would absolute values when
the column density becomes considerable ($> 10^{20}$~cm$^{-2}$). The
influence of such low-energy calibration errors on all other spectral
parameters is negligible.

\section{Spectral evolution of the flaring emission}
\label{sec:sevol}

For the purpose of determining the time evolution of the spectral
parameters of the flaring plasma, the individual time-resolved joint
LECS/MECS spectra have been fitted with with a one-temperature absorbed
thermal model with freely varying coronal abundance, plus a
two-temperature model with fixed spectral parameters, set to the
average values determined for the quiescent spectral emission. The
one-temperature model being fit thus represents the flare emission as
visible on top of the quiescent spectrum. The resulting best-fit
parameters for the flaring component are reported in
Table~\ref{tab:pars}.

\begin{figure*}[htbp]
  \begin{center}
    \leavevmode \epsfig{file=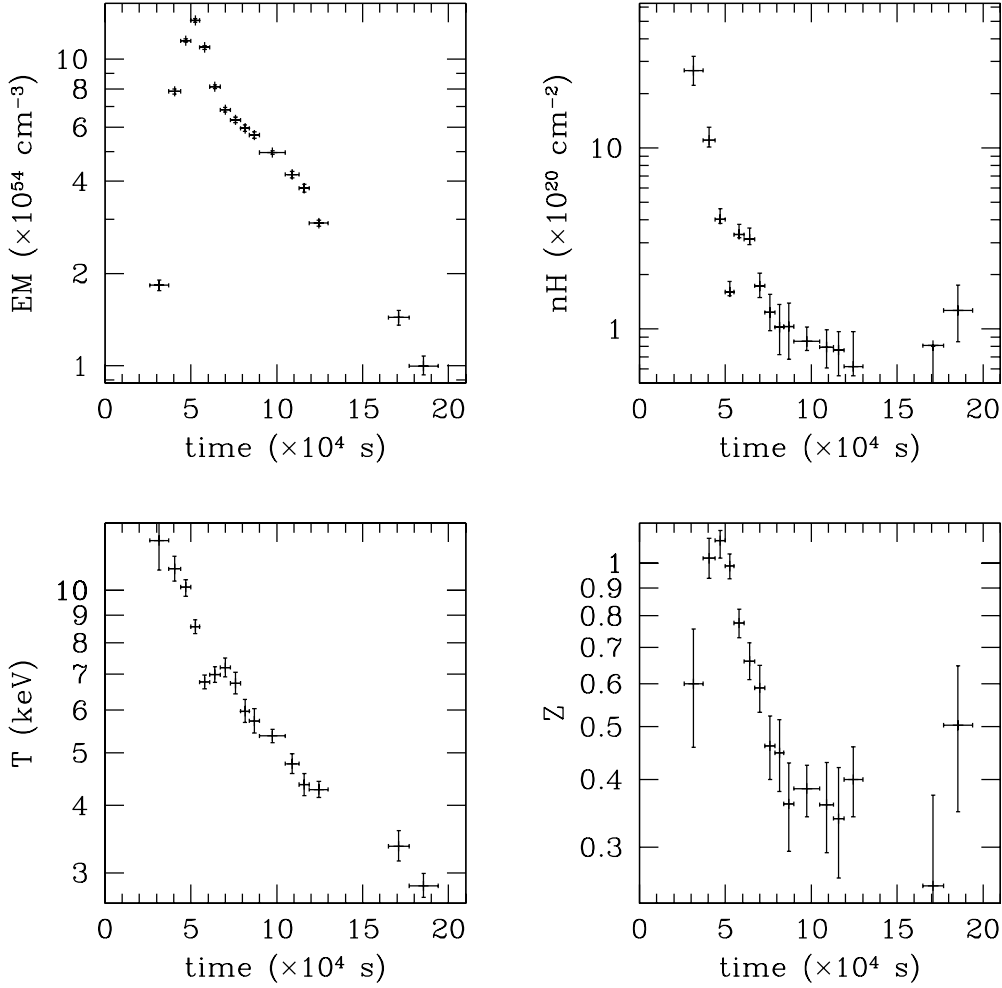, width=15cm, bbllx=10pt,
      bblly=150pt, bburx=600pt, bbury=700pt, clip=}
    \caption{The time evolution of the best-fit emission
      measure, temperature, metal abundance and absorbing column
      density for the best-fit parameters to the one-temperature flare
      spectra. Time is in seconds since MJD 51\,055 03:09:16. The
      best-fit parameters have been determined from the beginning of
      the large flare (time interval 2) until the beginning of the
      second flare visible in the observation (time interval 22). The
      time intervals affected by the eclipse of the flaring plasma
      (i.e.\ from 16 to 20) have also been excluded. The horizontal
      error bars represent the extent of the time intervals in which
      individual spectra have been accumulated, while the vertical
      error bars have been computed using the criterion $\Delta \chi^2
      = 3.50$, corresponding to the 68\% confidence level in the case
      of three interesting parameters. }
    \label{fig:tflare}
  \end{center}
\end{figure*}

The temporal evolution of the best-fit temperature, emission measure,
abundance and absorbing column density for the flaring component is
illustrated in Fig.~\ref{fig:tflare}. The one-temperature fits to the
flaring plasma emission provide a good description (i.e.\ a \cq\ value
close to 1.0) to most of the spectra, with the exception of the
spectra accumulated during the time-intervals 5 and 6, which yield a
reduced \cq\ which is formally unacceptable, corresponding to a very
low probability level. All other fits yield a reduced \cq\ 
corresponding to a probability that the adopted model gives a
satisfactory description of the data of $\ge 10$\%. We tried to
improve the quality of the fit to the spectra accumulated during time
intervals 5 and 6 by adding an additional thermal component to the
flare spectrum, i.e.\ assuming that the flare spectrum is better
described by a two-temperature model. While the additional degrees of
freedom lead to an improved \cq\ for flare spectra 5 and 6 when a
two-temperature model is used, the implied probability level for the
fit is still very low, and the two-temperature model is still not a
good description of the data. In addition, if a two-temperature model
is fitted to the flaring spectra for time intervals other than 5 and
6, the resulting spectral parameters are much more poorly determined,
with a high degree of degeneracy apparent if the confidence regions
are examined.

One obvious feature in the fit residuals for intervals 5 and 6 (which
is not altered by the use of a two-temperature model) is the bump at
$\simeq 1.2$~keV. A similar excess in the spectrum is visible, for
example, in the ASCA SIS spectrum of Capella discussed by
\cite*{bri98} and \cite*{bde+99}, which they attribute to a complex of
lines from Fe\,{\sc xvii} to Fe\,{\sc xix} from atomic levels with
high quantum numbers ($n>5$). These lines are missing in current
models, which will then consistently under-predict the emission in
this region, and are likely contributing to the higher \cq\ values
found for these time intervals.  While one possible way of decreasing
the residuals would be to allow selected abundance ratios to vary
during the fit, this would not however be granted for the rest of the
flare spectra, and doing it only for intervals 5 and 6 would again
yield results which cannot be compared with the rest.  Also, as shown
by \cite*{bde+99}, the lack of the high-excitation Fe lines in the
model yields spurious abundances for the other elements if they are
left free to vary; we will therefore not explore this possibility
further in the present paper.

\begin{figure*}[htbp]
  \begin{center}
    \leavevmode \epsfig{file=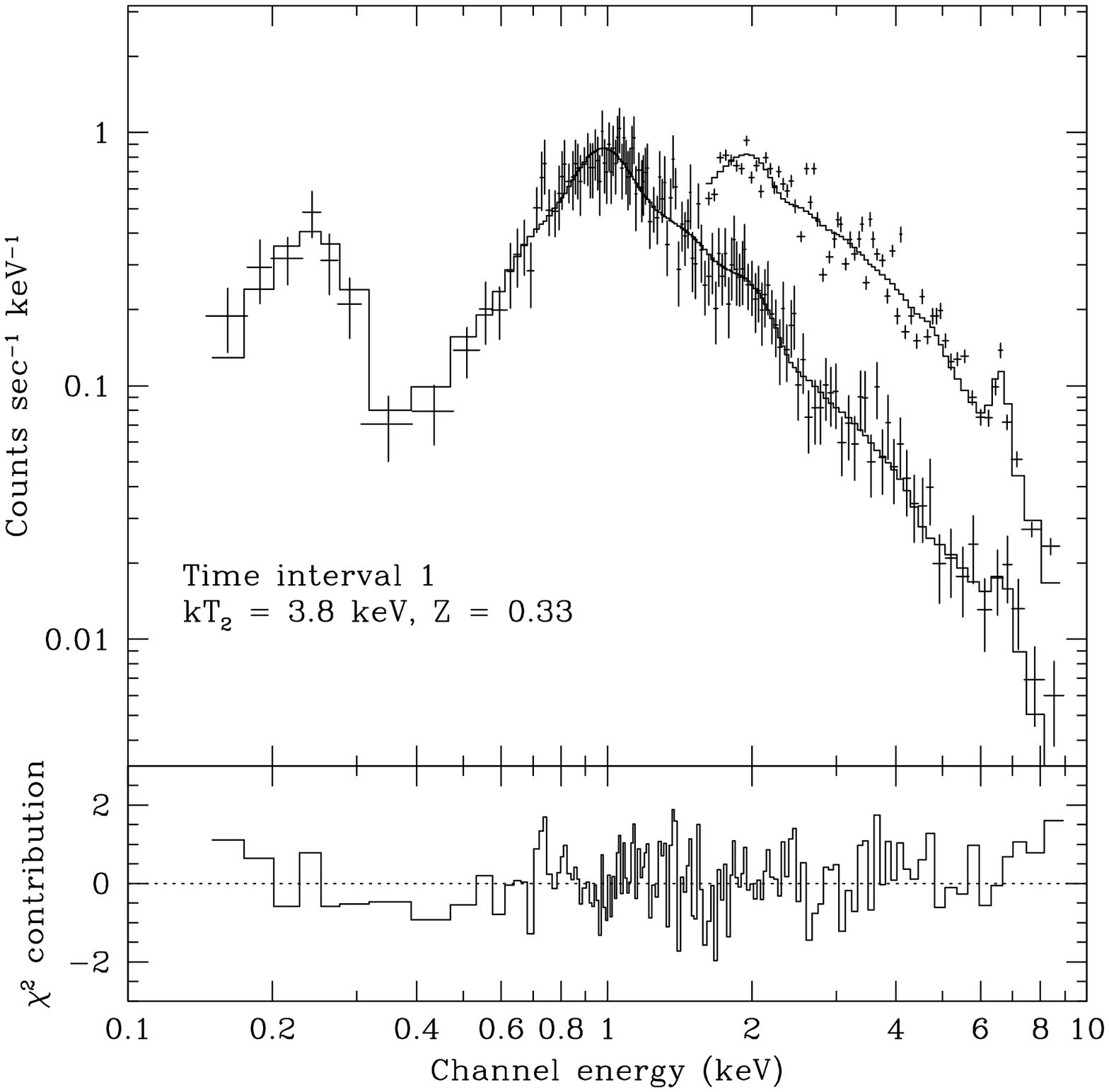, width=5.8cm, bbllx=20pt,
      bblly=150pt, bburx=600pt, bbury=700pt, clip=}
    \leavevmode \epsfig{file=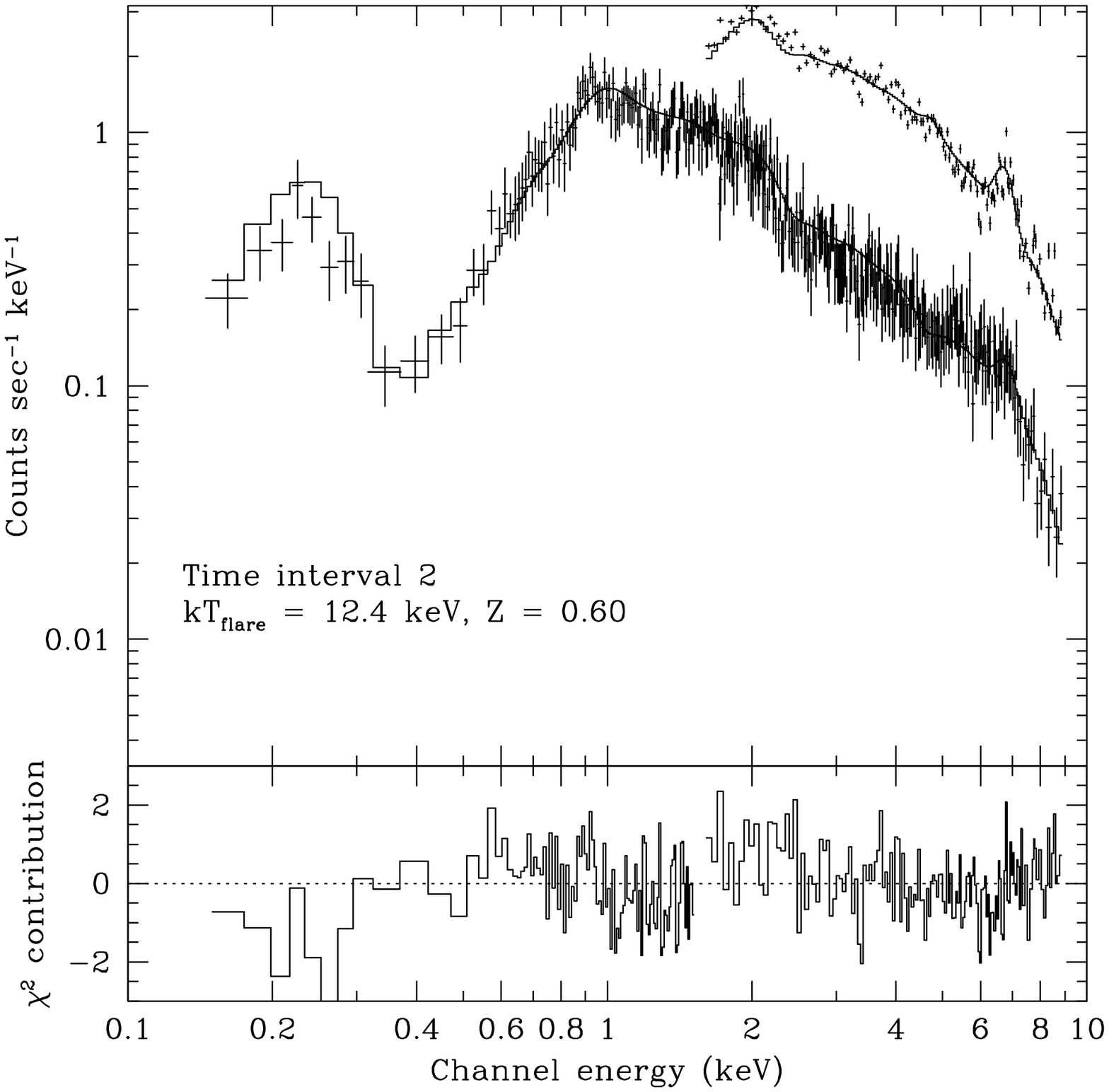, width=5.8cm, bbllx=20pt,
      bblly=150pt, bburx=600pt, bbury=700pt, clip=}
    \leavevmode \epsfig{file=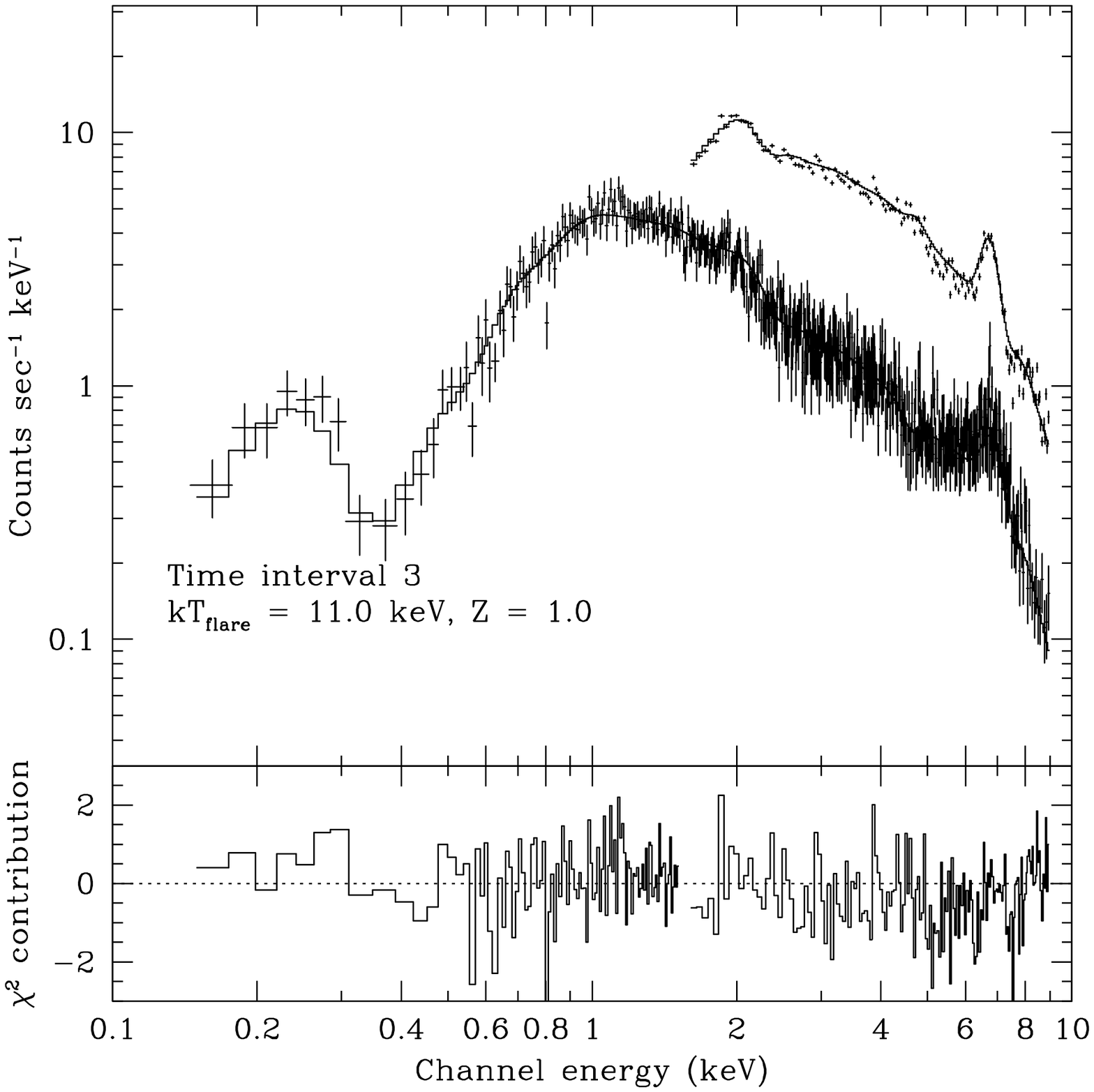, width=5.8cm, bbllx=20pt,
      bblly=150pt, bburx=600pt, bbury=700pt, clip=}
    \leavevmode \epsfig{file=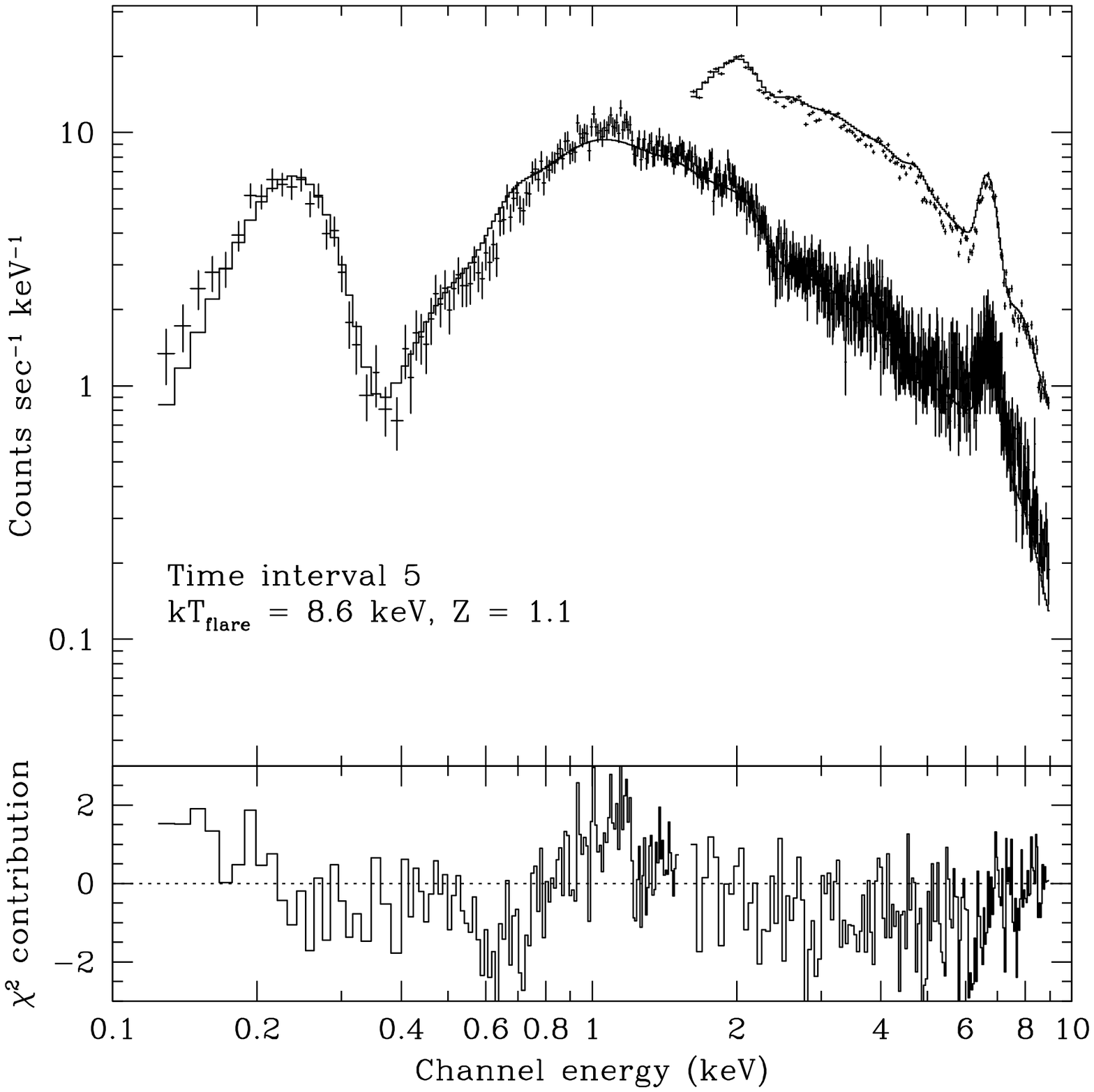, width=5.8cm, bbllx=20pt,
      bblly=150pt, bburx=600pt, bbury=700pt, clip=}
    \leavevmode \epsfig{file=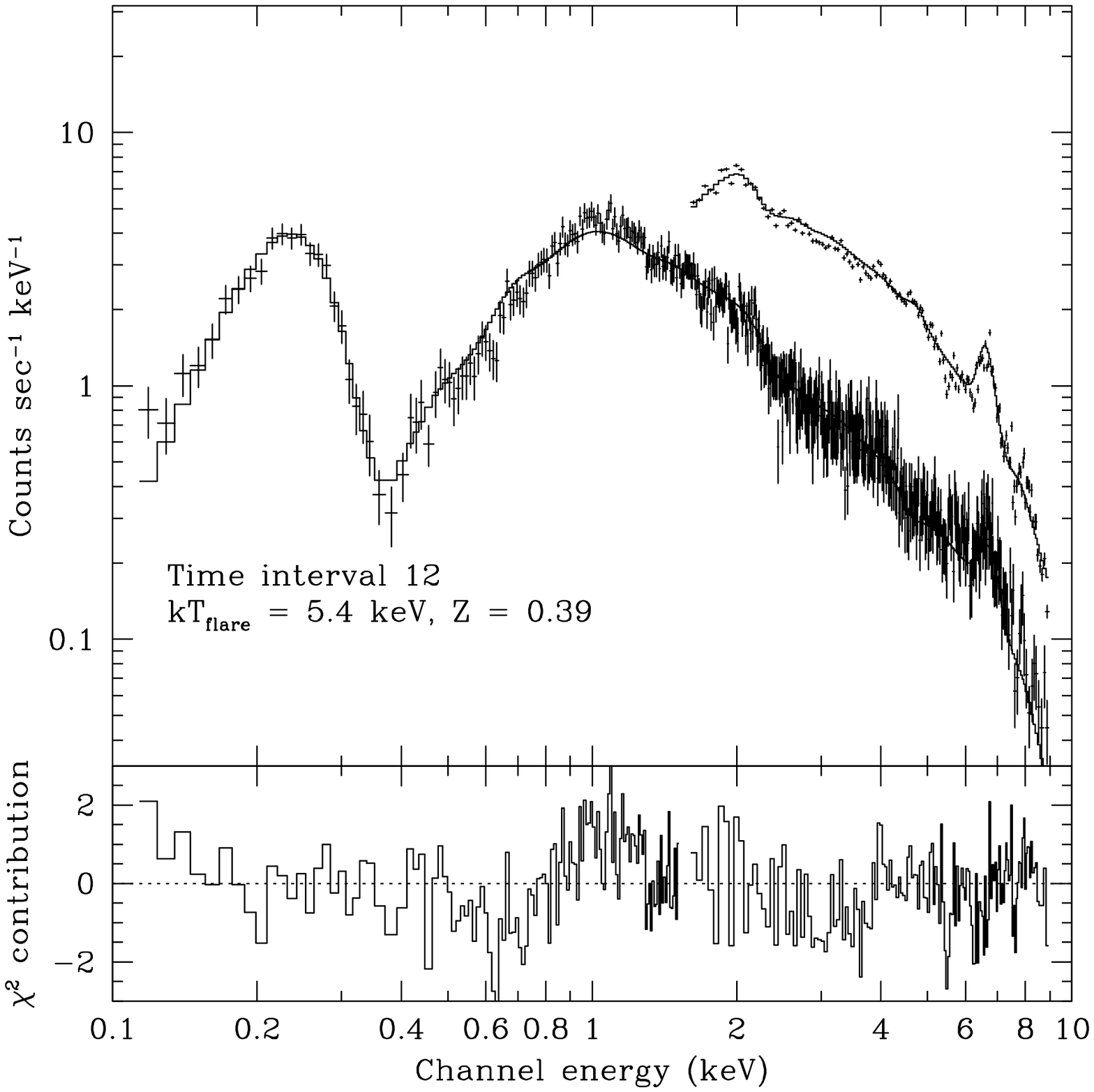, width=5.8cm, bbllx=20pt,
      bblly=150pt, bburx=600pt, bbury=700pt, clip=}
    \leavevmode \epsfig{file=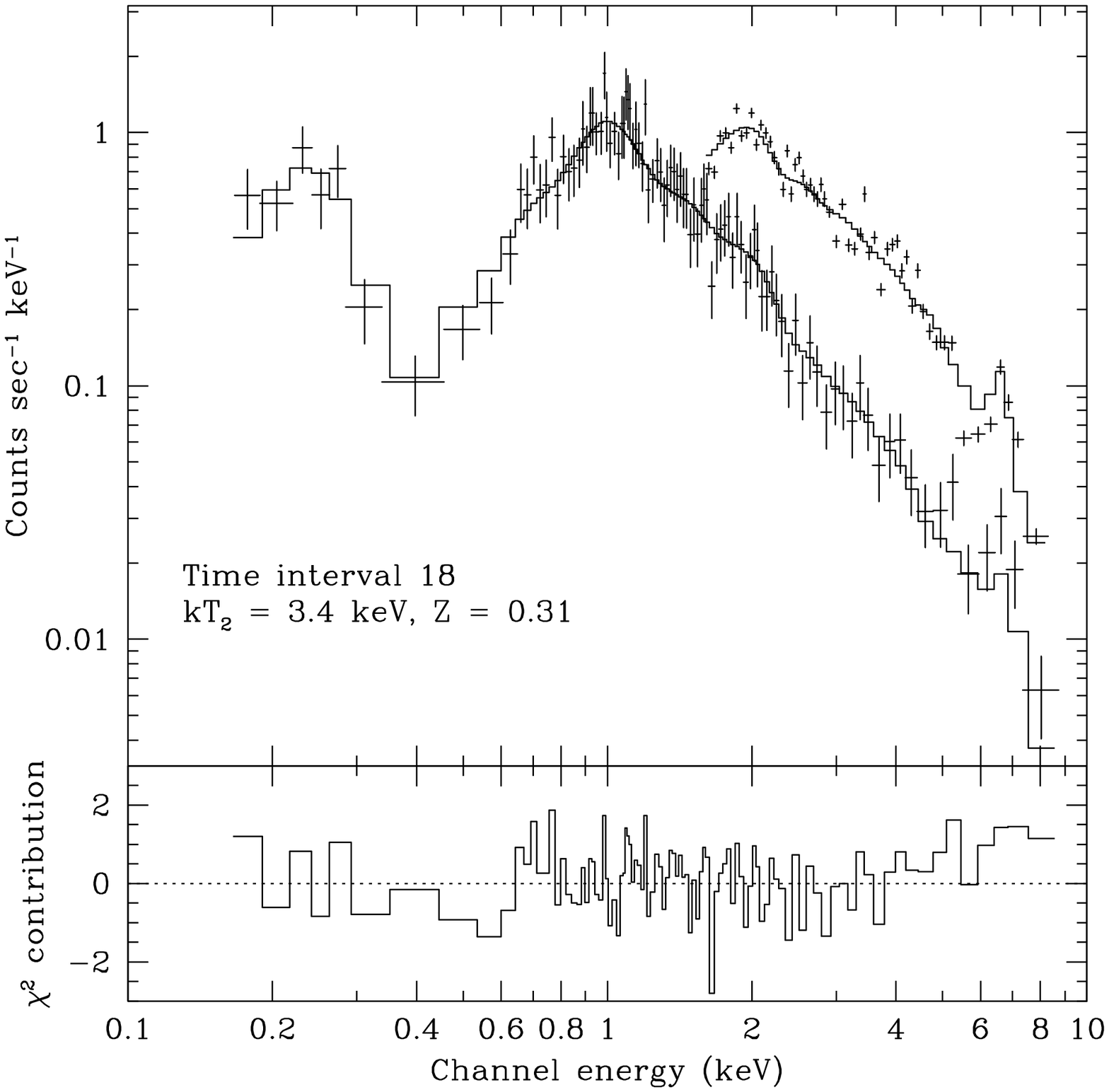, width=5.8cm, bbllx=20pt,
      bblly=150pt, bburx=600pt, bbury=700pt, clip=}
    \caption{The LECS and MECS-2 spectra recorded during the time
      intervals 1, 2, 3, 5, 12 and 18 (as indicated in
      Fig.~\ref{fig:lc}). The MECS-2 spectra have been shifted upwards
      by a factor of 5 for clarity. Each plot shows the observed
      spectra, the best fit model, and, in the lower panel, the
      (signed) contribution to the \cq\ of each bin. The \cq\ 
      contributions are plotted for the LECS spectrum for $E<1.6$\,keV
      and for the MECS for $E>1.6$\,keV. The best-fit model plotted is
      a two-temperature absorbed {\sc mekal} model with varying
      abundance for intervals 1 and 18, while for the flare spectra
      (intervals 2, 3, 5 and 12) a one-temperature component with
      varying abundance has been added on top of the quiescent
      emission, as described in the text. All the spectra shown have
      been rebinned at 20 channels per (variable size) bin. Note the
      change in the vertical scale for the spectra from time segments
      3, 5 and 12.}
    \label{fig:espec}
  \end{center}
\end{figure*}

From Fig.~\ref{fig:tflare} the good temporal coverage of the evolution
of the flare is evident, from its onset all the way to its
disappereance. Note the gap between $\simeq 130$ and $\simeq 160$~ks
due to the eclipse. The general behavior of the flare is similar to
the one commonly observed in the several large stellar and solar
flares studied so far, with the temperature peaking at the beginning
of the flare and the emission measure rising more slowly and peaking
at a later time. The shape of the temperature decay is close to
exponential throughout the flare (although it briefly increases again
between $\simeq 50$ and $\simeq 70$~ks). The emission measure
increases slowly for a long time ($\sim 20$\,ks) after the temperature
has peaked, and its decay is not well described by a single
exponential, with a more rapid decay observable in the first $\sim
20$\,ks of the flare, and a longer decay time-scale observable
afterwards, closely mirroring the behavior of the 1.6--10~keV light
curve.

\subsection{Abundance variations during the flare}

As discussed in Sect.~\ref{sec:intro}, previous observations of strong
flares on Algol performed with the GINGA and ROSAT observatories
hinted at variations of the coronal metal abundance during the flaring
event. However, the limited spectral coverage and resolution of the
GINGA and ROSAT proportional counters made it difficult to fully
disentangle abundance effects from other effects, such as changes in
the absorbing column density for the PSPC or changes in the
temperature structure in the case of GINGA. The combination of
resolution and spectral coverage of the LECS and MECS detectors allows
to effectively disentangle the effects of the plasma metal abundance
on the emitted spectrum from the thermal structure (for a discussion
see \cite{fmp+97}, b).  This, coupled with the excellent time coverage
of the BeppoSAX observation and the slow flare decay, allows to study
in detail the evolution of the coronal metal abundance during the
flare.

The abundance of the quiescent plasma (i.e. the best-fit value of the
two-temperature model to the spectra accumulated during the intervals
0, 1 and 18) is $\simeq 0.3$ times the solar photospheric one, a value
compatible with the abundance derived for the quiescent Algol corona
by \cite*{sls+95} on the basis of an analysis of the EUVE spectrum.
The temporal evolution of the best-fit abundance of the flaring plasma
is shown in Fig.~\ref{fig:tflare}.  Consistent with the indications of
the ROSAT and GINGA data, the metal abundance of the flaring plasma
increases significantly during the early phases of the flare, to a
value of approximately 1.0, and then rapidly decays back to a value
consistent with the one determined from the pre-flare spectrum.

The time scale for the increase of the coronal abundance is similar to
the time scale with which the emission measure increases, while the
abundance ``decay'' time is significantly faster than either the flare
temperature or the emission measure decay times, so that the coronal
abundance goes back to its pre-flare value while a significant flaring
component is still well visible in the spectrum (with significant
excess emission measure and a temperature of some $\sim 4$\,keV). The
decaying part of the time evolution of the best-fit abundance is well
fit with a single exponential, with an $e$-folding time of 36~\,ks,
with the further evolution compatible with a constant abundance.

The effect of the varying abundance in the flaring component is
clearly visible in the set of spectra plotted in Fig.~\ref{fig:espec}.
The pre-flare spectrum (from segment 1) is well fit by a
two-temperature plasma with abundance $Z \simeq 0.3$, and a maximum
temperature $T \simeq 4$\,keV. The spectrum immediately following
(from segment 2) marks the beginning of the flare, and shows the
presence of the very hot flaring component (with $T \simeq 12$\,keV),
although still with relatively little emission measure and only a
small enhancement ($\le 2$ times) in the plasma abundance. In the
subsequent spectrum (from segment 3) the temperature has started
decaying, while the emission measure is still rising.  The plasma
abundance has reached its peak at $Z \simeq 1$, so that the Fe\,K
complex is well visible.  The spectrum from segment 5 is
characteristic of the flare's emission measure peak. The temperature
has decreased to $\simeq 8.6$\,keV, while the abundance is still close
to the solar value. In the spectrum from segment 12 the emission
measure has decayed to less than half the peak value, the flare
temperature is down to $\simeq 5.4$\,keV and the abundance is back to
almost the quiescent value ($\simeq 0.35$), as again evident in the
near disappearance of Fe\,K complex.

Could the changes in the best-fit plasma abundance parameter be
explained through mechanisms other than a physical change in the
abundance? With the exception of two of the time intervals discussed
above, the model used yields fully acceptable reduced \cq\ values,
thus making it not necessary to invoke, from a statistical point of
view, any other mechanism. For the flare spectra the fit process is
driven, in the determination of the abundance, essentially by one
diagnostic, i.e. the intensity of the Fe\,K line complex.  For most
coronal sources the intensity of the lines in the rich Fe\,L line
complex will be a more important diagnostic of metallicity, given the
much higher signal-to-noise ratio and thus statistical weight of the
lines in this region of the spectrum, where both the intrinsic photon
flux and the instrument's effective area are higher. However, when the
emission is dominated by a plasma at temperatures higher than $\simeq
5 $\,keV (as in the case of the Algol's flare spectrum), the spectrum
shows very little line emission (specially from Fe\,L lines), and the
only prominent line is the one due to the Fe\,K complex.  Therefore,
changing the abundance in the flaring component during the hot phase
of the flare leaves the model spectrum essentially unchanged except
for the intensity of the Fe\,K complex.

\begin{figure}[htbp]
  \begin{center}
    \leavevmode \epsfig{file=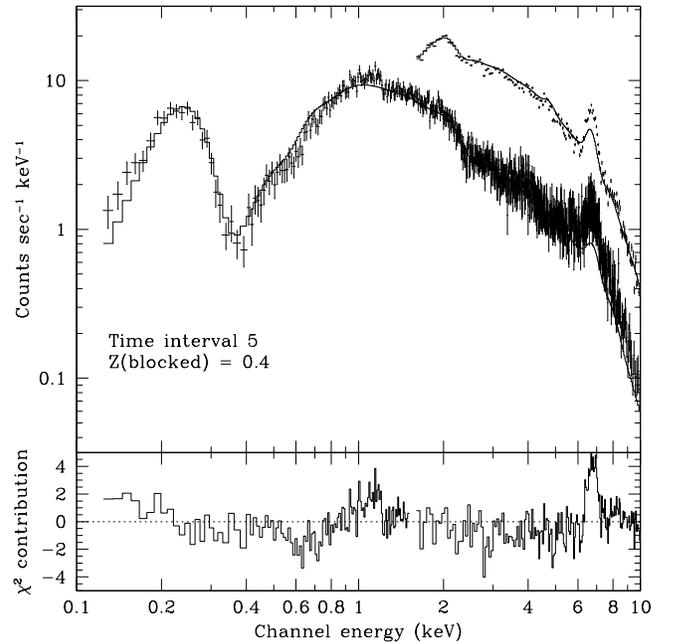, width=9.5cm, bbllx=10pt,
      bblly=150pt, bburx=600pt, bbury=700pt, clip=}
    \caption{The spectrum accumulated during interval 5, shown here
      with a best-fit model for the flaring emission obtained with a
      single thermal component and a plasma abundance fixed at $Z =
      0.4 \times \zsun$.  Channels from 5.5 to 8.0\,keV were excluded
      from the fit, to remove the influence of the Fe\,K complex
      emission. They are however plotted here to show the difference
      in the predicted strength of the Fe\,K complex.}
    \label{fig:t5lowz}
  \end{center}
\end{figure}

To show this, we have fit the spectrum from interval 5 with the same
type of model (one-temperature on top of the quiescent emission), but
excluding from the fit the Fe\,K region (i.e.\ channel energies from
5.5 to 8.0\,keV) and with the abundance of the flaring component fixed
to $Z = 0.4 \times \zsun$ (i.e.\ compatible with both the quiescent
abundance value and the late-decay flaring value). The resulting
best-fit model is shown in Fig.~\ref{fig:t5lowz}, in which the Fe\,K
region has also been plotted. Inspection of the residuals (and their
comparison with the relevant panel of Fig.~\ref{fig:espec}) shows that
$Z = 0.4 \times \zsun$ model provides a good fit to all of the
spectrum (with the residuals showing the same structure and size as in
the case of the $Z \simeq 1.0~\zsun$ model), but will, as expected,
strongly under-predict the Fe\,K complex. As evident from the size of
the residuals at the position of the Fe\,K complex, they are driving
the abundance determination in the flaring emission. Restricting the
fit to a higher-energy interval, e.g.\ to the spectral region harder
than 3~keV, does not change the best-fit abundance, although the
resulting confidence region are of course broader.

Fluorescence from the heated photosphere has been proposed as a
possible mechanism for the enhancement of the Fe\,K emission during a
strong flare. In this case, however, the emission would come from low
ionization states of Fe, and the line energy would thus be
significantly different ($E \simeq 6.4$~keV). At the resolution of the
MECS detectors, and at the signal-to-noise ratio of the spectra
discussed here, such large shift in energy of the line centroid would
easily be seen. We have determined the 90\% upper limit to the
intensity of a 6.4~keV Fe\,K line to be $\le 10^{-3}$ the intensity of
the observed 6.8~keV line complex, so that this explanation can be
ruled out.  The same lack of shift in the line energy allows to rule
out significant non-equilibrium ionization effects. Also, the
equilibrium time scale for a plasma of this high temperature is of
order of few tens of seconds, negligible in comparison with the long
time scales on which the abundance is observed to vary.

Given the relative simplicity of the formation mechanism of the Fe\,K
line complex, as well as the absence of other contaminating features
in the spectrum, other explanations of the observed strong increase of Fe\,K
emission than an actual abundance increase in the emitting
plasma appear rather difficult to find.

\begin{figure*}[htbp]
  \begin{center}
    \leavevmode \epsfig{file=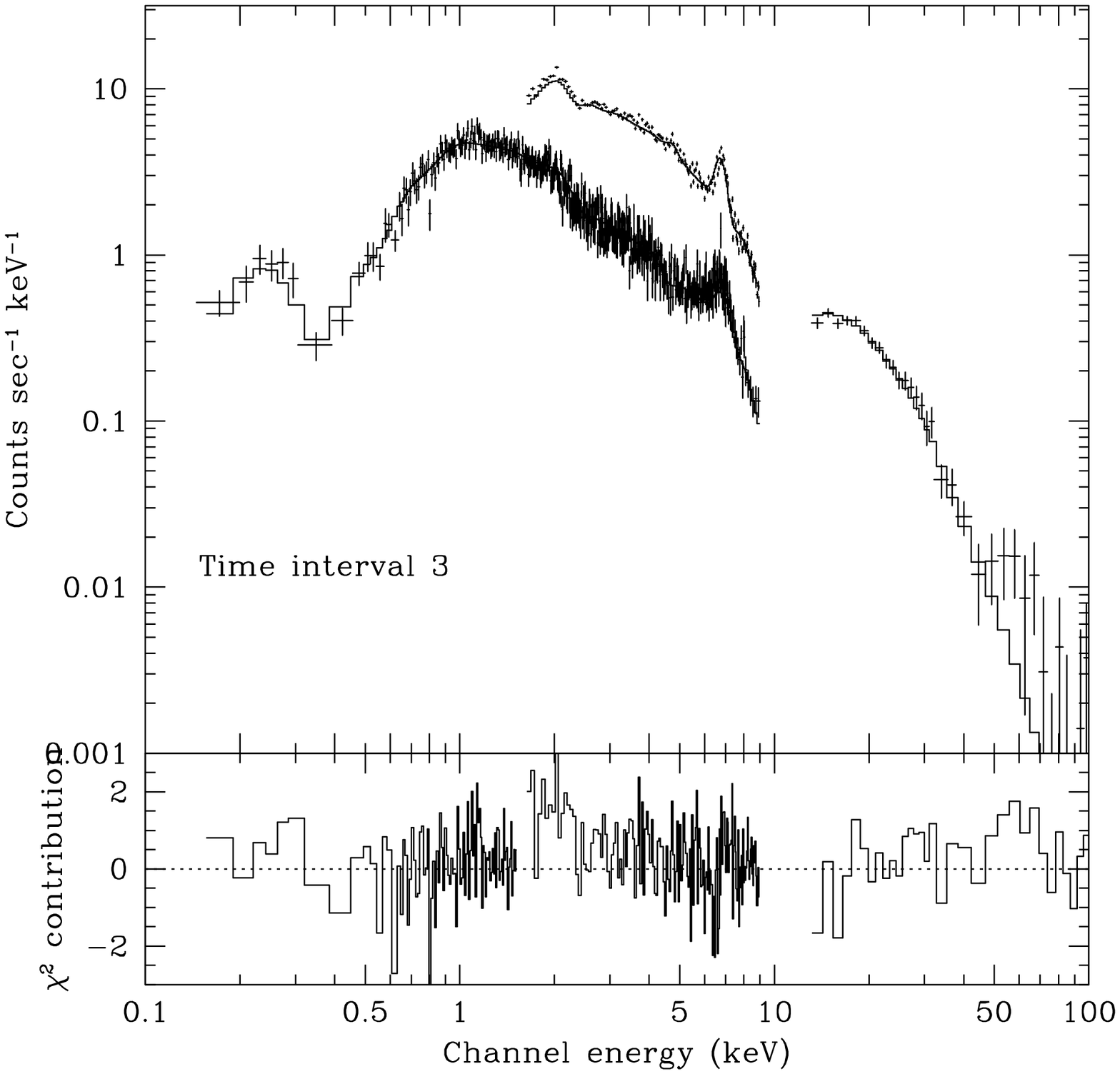, width=5.8cm, bbllx=20pt,
      bblly=150pt, bburx=600pt, bbury=700pt, clip=}
    \leavevmode \epsfig{file=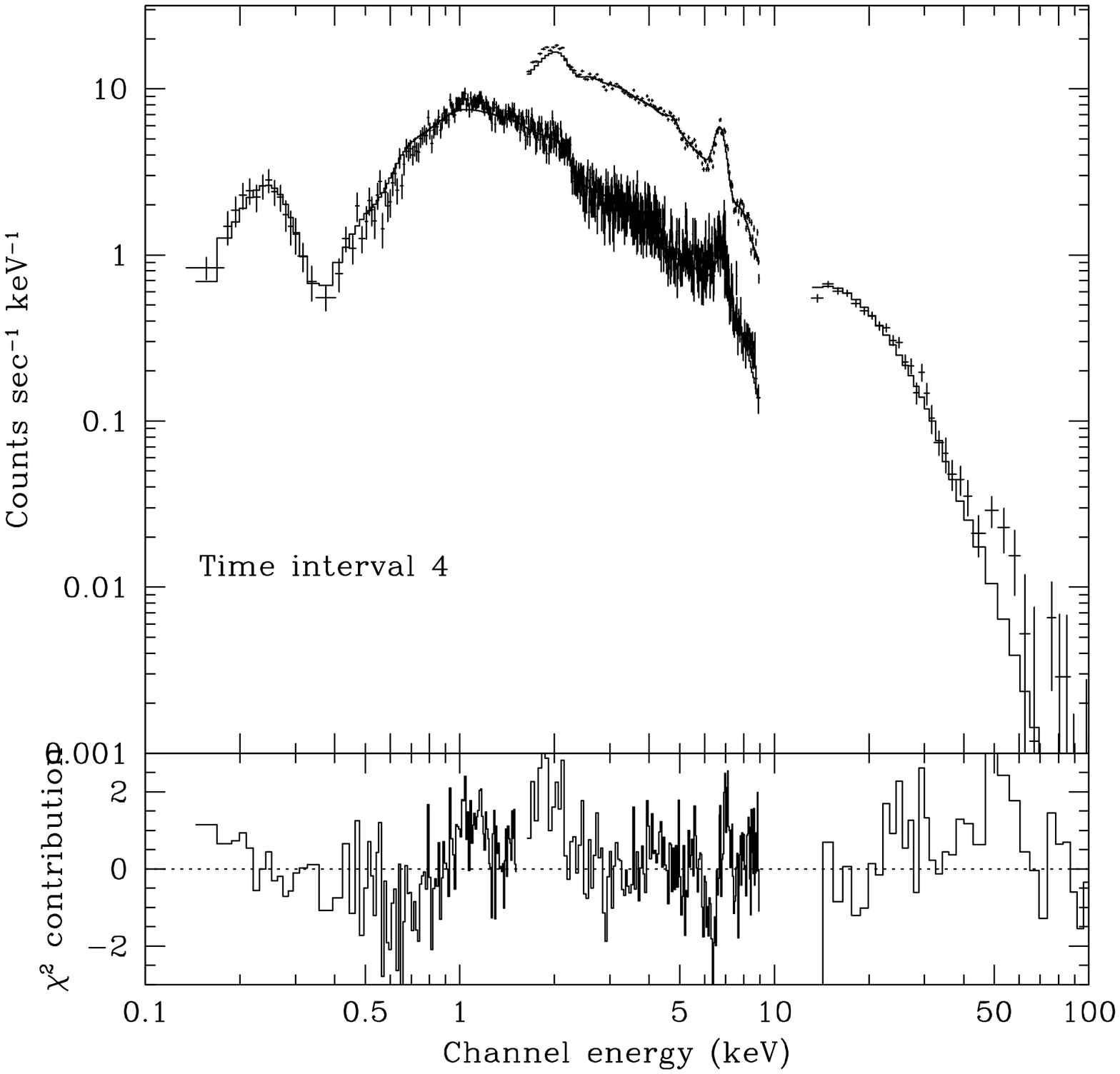, width=5.8cm, bbllx=20pt,
      bblly=150pt, bburx=600pt, bbury=700pt, clip=}
    \leavevmode \epsfig{file=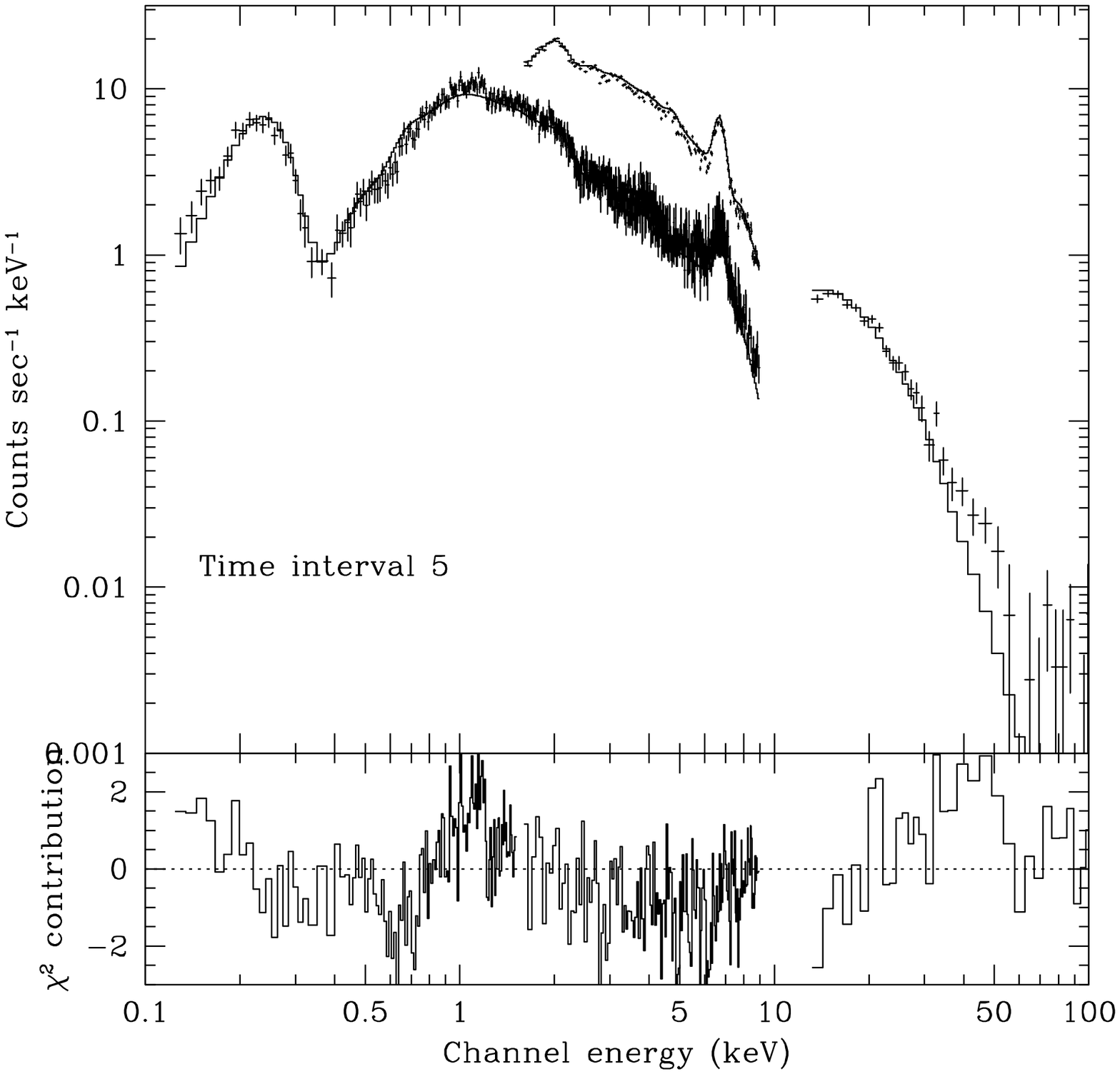, width=5.8cm, bbllx=20pt,
      bblly=150pt, bburx=600pt, bbury=700pt, clip=}
    \leavevmode \epsfig{file=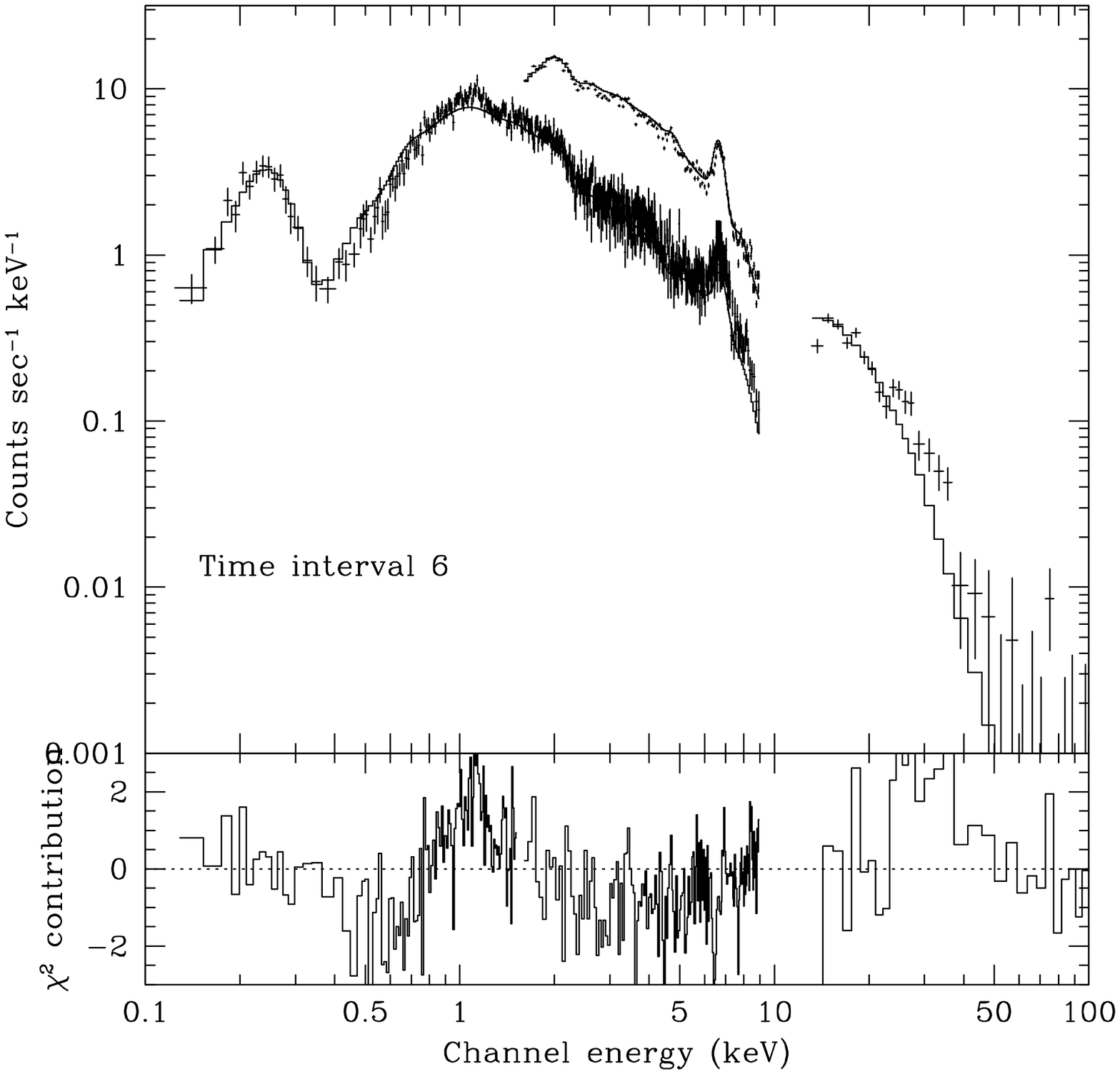, width=5.8cm, bbllx=20pt,
      bblly=150pt, bburx=600pt, bbury=700pt, clip=}
    \leavevmode \epsfig{file=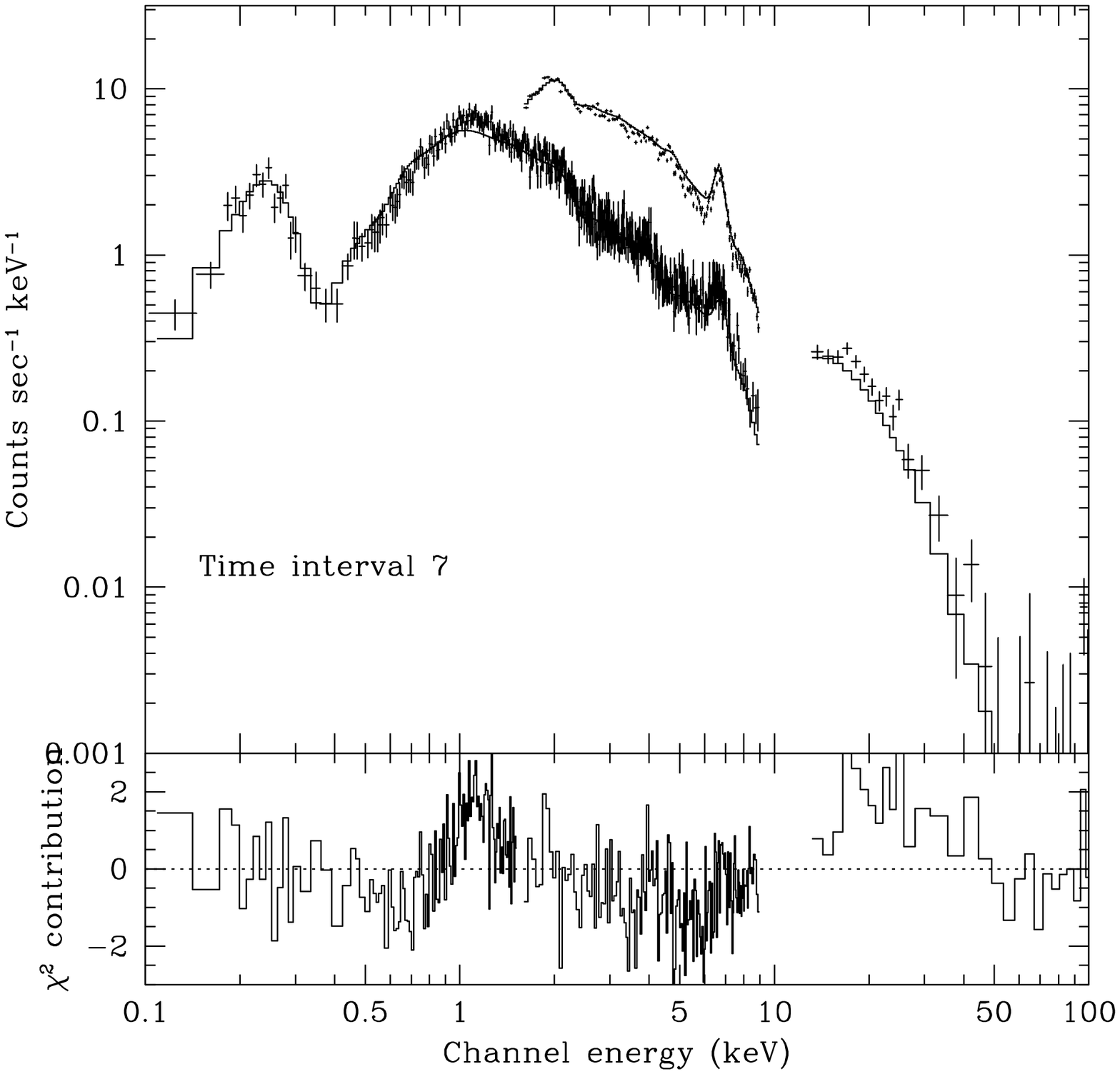, width=5.8cm, bbllx=20pt,
      bblly=150pt, bburx=600pt, bbury=700pt, clip=}
    \caption{The LECS, MECS-2 and PDS spectra recorded during the time
      intervals 3 to 7 (as indicated in Fig.~\ref{fig:lc}). The MECS-2
      spectra have been shifted upwards by a factor of 5 for clarity.
      Each plot shows the observed spectra, the best fit model, and,
      in the lower panel, the (signed) contribution to the \cq\ of
      each bin. The \cq\ contributions are plotted for the LECS
      spectrum for $E<1.6$\,keV, for the MECS for $1.6\,{\rm keV} < E
      < 10$\,keV and for the PDS for $13\,{\rm keV} < E < 100$\,keV.
      The best-fit model plotted is a one-temperature component with
      varying abundance on top of a two-temperature absorbed {\sc
        mekal} model describing the quiescent emission, as described
      in the text.}
    \label{fig:pspec}
  \end{center}
\end{figure*}

\subsection{Absorbing material}
\label{sec:abs}

Inspection of Table~\ref{tab:pars} shows that the best-fit value for
the absorbing column density during the initial phase of the flare is
very high ($N({\rm H}) \ga 10^{21}$\,cm$^{-2}$). This high best-fit
absorbing column density is clearly driven by the depressed soft
continuum of the initial flare spectra. A comparison of the spectra
observed during time intervals 1 (pre-flare) and 2 and 3 (beginning of
the flare) as plotted in Fig.~\ref{fig:espec} shows that while the
emission above $\simeq 1$\,keV increases considerably once the flare
begins, the emission in the region below $\simeq 0.5$\,keV is
essentially unchanged from its pre-flare values. Given the high
temperature of the flaring component ($T \simeq 10$\,keV) in this
phase (which implies an essentially featureless continuum for the
flaring spectrum, with the exception of the Fe\,K complex emission),
the only way to obtain a spectrum significantly depressed in the
softer region is by introducing an absorbing column density for the
flaring component only.  Indeed, inspection of the best-fit model
spectrum shows that during time intervals 2 and 3 the flaring spectrum
is completely absorbed below $\simeq 0.5$\,keV, so that the emission
visible is only due to the quiescent component (for which the
absorbing column density is fixed). Later during the flare evolution
the absorbing column density decreases, so that the soft emission (as
visible for time interval 5 in Fig.~\ref{fig:espec}) increases by a
factor of $\simeq 10$, to an intensity proportional to the harder part
of the spectrum.

The change in absorbing-column density during the flare evolution is
also visible in the light curves in the soft (0.1--0.5~keV) and hard
(1.6--10.0~keV) band (i.e.\ Figs.~\ref{fig:lc} and~\ref{fig:lcle}): in
the soft band the flare only begins in interval~3, while in the hard
band the count rate has already risen by a factor of 10 at the end of
interval~2. This apparent ``delay'' in the flare onset in the softer
band can be explained by the high best-fit absorbing column density.
The soft-band light-curve also shows a much slower decay than the
hard-band one, with an actual rate increase between intervals~6 and~8,
and an essentially flat behavior between intervals 8 and 12.

Is the best-fit absorbing column density actually due to real material
in the line of sight? An obvious alternative explanation would be that
the fit process uses the absorbing column density to compensate for
deficiencies in the plasma emission codes. However, given the
simplicity of a thermal spectrum at the high temperature observed at
the beginning of the flare (dominated by continuum emission) it is
hard to imagine that the softer part of the spectrum could be
under-predicted by a factor of ten. We thus consider the presence of
local absorbing material, depressing the soft emission at the
beginning of the flare, to be the correct explanation.

One possible interpretation for this is in terms of a massive coronal
mass ejection taking place at the beginning of the flare, which
provides the absorbing material. 

\subsection{The PDS spectra}

The PDS detector on-board the BeppoSAX observatory is sensitive to
X-rays approximately in the passband from $\simeq 15$ up to $\simeq
300$\,keV.  Most coronal sources have too little flux in this band to
be detected. However, given its very high temperature, the emission
from the Algol flare produces an easily detectable signal also in the
PDS.  We extracted individual PDS spectra from the time intervals 3 to
7 (inclusive), using the tools provided within the {\sc saxdas}
package, and jointly fitted the LECS, MECS and PDS data for these
intervals, with the same model (a one temperature model superimposed
with a two-temperature quiescent emission) which we used for the
analysis of the LECS and MECS only. Unfortunately, while it would have
been desirable to include the PDS data in the analysis of the time
dependence of the flare parameters, the temporal coverage afforded by
the PDS data is more limited, and for most of the decay phase no PDS
spectra with sufficient signal to noise are available.  Thus, to keep
the set of parameters for the flare decay homogeneous, we have only
used the LECS and MECS data in that context.

The sequence of time-resolved PDS spectra is shown, together with the
best-fit model, in Fig.~\ref{fig:pspec}. One of the main questions for
which the PDS data are of interest is whether any non-thermal spectral
components are present in the flare spectrum. The spectra of
Fig.~\ref{fig:pspec}, together with their residuals, show that the PDS
spectrum of flare peak is well described as the high-energy tail of
the hot flaring plasma and no additional non-thermal components appear
to be present.

All of the PDS spectra present a ``bump'' in the region close to
$\simeq 50$--60\,keV (particularly visible in the spectrum from
interval 4), plus a second minor bump close to 30\,keV. Such features
are, according to PDS calibration team (D.~Dal Fiume, private
communication) of instrumental origins, due to a blend of residual
lines from the Am calibration source with instrumental fluorescence
features due to the Ta collimator (feature at 50--60\,keV), while the
feature at 30\,keV is the escape peak of the feature at higher energy.

The temperature of the flaring component derived from the joint LECS,
MECS and PDS fit is, within $1\,\sigma$, the same as the temperature
derived from the LECS and MECS data only, thus confirming that the
spectrum seen in the PDS detector is the tail of the thermal spectrum
seen in the LECS and MECS passbands.

\subsection{Energetics}
\label{sec:energy}

\begin{figure}[htbp]
  \begin{center}
    \leavevmode \epsfig{file=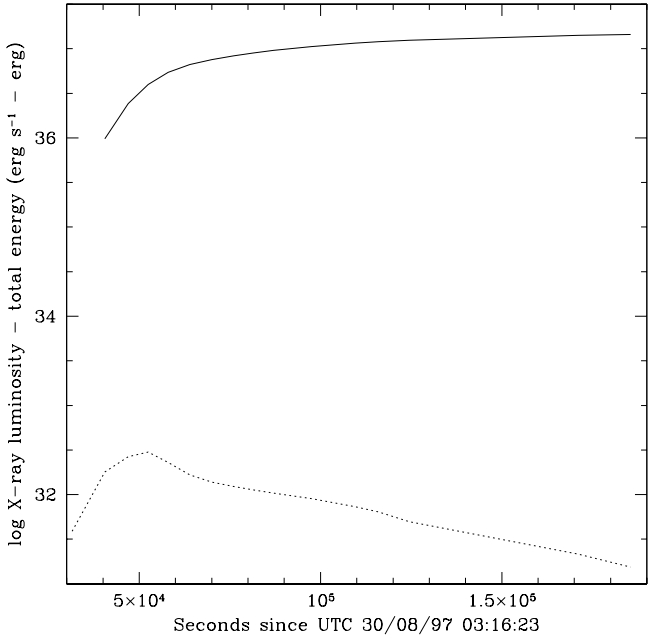, width=9.5cm, bbllx=5pt,
      bblly=150pt, bburx=600pt, bbury=700pt, clip=}
    \caption{The evolution of the flare's X-ray luminosity in the
      0.1--10~keV band (lower curve) together with the total X-ray
      emitted energy in the same band (higher curve).}
    \label{fig:energy}
  \end{center}
\end{figure}

From the detailed temporal evolution of the flare's spectral parameter
it is possible to derive the instantaneous luminosity in the
0.1--10~keV band and therefore, by integration, the total energy
emitted in soft X-rays during the flare. The temporal evolution of the
X-ray luminosity is shown in Fig.~\ref{fig:energy}.

The total X-ray radiative loss at the end of the flare is $1.4 \times
10^{37}$~erg, making this one of the most energetic X-ray flares ever
observed on stellar sources, on a par with the long flare observed on
CF~Tuc by \cite*{ks96}.

Given the short thermodynamic decay time of the loop implied by the
eclipse-derived size ($\la 10$~ks, see Sect.~\ref{sec:length}), the
loop has, on the total time scale of the flare, negligible thermal
inertia, i.e.\ it will quickly respond to changes in the heating.
Thus, the observed X-ray luminosity temporal evolution closely
describes the temporal evolution of the heating, slowly rising for
some tens of ks and then decaying for more than 100~ks.  Whatever the
mechanism ultimately responsible for the heating is, it thus must be
capable of operating on these time scales.

\section{Analysis of the flare decay}

Different approaches have been proposed to determine, from the
temporal evolution of the flare temperature and emission measure, the
parameters, and in particular the size (and thus density) of the
flaring region. One unique feature of the Algol flare discussed here
is the availability of an actual measurement of the physical size of
the flaring region from the duration and shape of its eclipse
(\cite{sf99}).  Thus, for the first time we can test whether the
different methods used to analyze the flare decay phase yield answers
consistent with the size of the flaring region implied by the eclipse.

\subsection{Assumptions}

All the methods discussed in the literature rely on the analysis of
the decay phase of the flare, defined as the phase of the flare during
which both the temperature and the total number of emitting particles
(using the emission measure, or more properly its square root, as
proxy) decrease. This is equivalent to saying that the total energy
content of the flaring loop is decreasing.

\subsection{The quasi-static cooling formalism}
\label{sec:qs}

In this Section we follow the quasi-static formalism for the analysis
of the decay phase of a flare, as developed by \cite*{om89} and as
applied, for example, to the analysis of the large PSPC flare on Algol
by \cite*{os96}. In this framework the information about the linear
size of the flaring loop is derived from the observed relaxation times
of the cooling plasma, under the assumption that the two relevant
mechanisms for energy loss from the flaring plasma are radiation and
thermal conduction along the loop. The ratio between the conductive
and radiative time scales ($\tau_{\rm c} / \tau_{\rm r}$) is in
general unknown, given that $\tau_{\rm c}$ will depend on the geometry
of the flaring loop(s).  The quasi-static formalism requires this
ratio to be constant. Under this condition the geometry of the loop
can then be determined.

Following \cite*{omb88}, the effective decay time for the thermal
energy during the loop decay is defined as
\begin{equation}
  \label{eq:taueff}
  {{1}\over{\tau_{\rm eff}}} = {{1}\over{\tau_{\rm r}}} + 
  {{1}\over{\tau_{\rm c}}} = {{(1+ \gamma/2)}\over{\tau_{T}}} +
  {{1}\over{2 \tau_{\rm d}}}
\end{equation}
where ${\tau_{T}}$ is the decay time of the loop temperature,
$\tau_{\rm d}$ is the decay time of the light-curve and $\gamma$ is
the power law exponent of the temperature dependence of the radiative
cooling function $\Psi$ of an optically thin plasma hot plasma. This
function is usually parameterized as $\Psi (T) \ = \ \Psi_0
T^{-\gamma}$, and for sufficiently hot plasmas ($T \ga 20$\,MK) the
values $\Psi_0 = 10^{-24.73}$ and $\gamma \simeq -0.25$ provide a very
good parameterization of the radiative cooling losses.  Note that at
these temperatures the emission is dominated by the continuum, with
very little line contribution, and hence the emissivity does not
depend on the metal abundance, making this approach applicable even in
the presence of observed abundance variations.

The radiative and conductive decay times are defined as
\begin{equation}
  \label{eq:taur}
  \tau_{\rm r} = {{3 n_0 k_{\rm B} T_0} \over {n_0^2 \Psi_0 T_0^\gamma}}
\end{equation}
and 
\begin{equation}
  \label{tauc}
  \tau_{\rm c} = {{3 n_0 k_{\rm B} T_0} \over 
    {8 \kappa_0 T_0^{7/2} f(\Gamma)} 
    (\Gamma + 1) L^2}
\end{equation}
where $n_0$ and $T_0$ are the electron density and the temperature at
the beginning of the decay phase and $L$ and $\Gamma$ are the length
and expansion factor of the loop. $f(\Gamma)$ is a correction factor
accounting for the change in the conductive flux for tapered loops.
The adopted value for the conductivity is $\kappa_0 = 8.8 \times
10^{-7}$\,erg\,cm$^{-1}$\,K$^{-7/2}$.

\begin{figure}[htbp]
  \begin{center}
    \leavevmode \epsfig{file=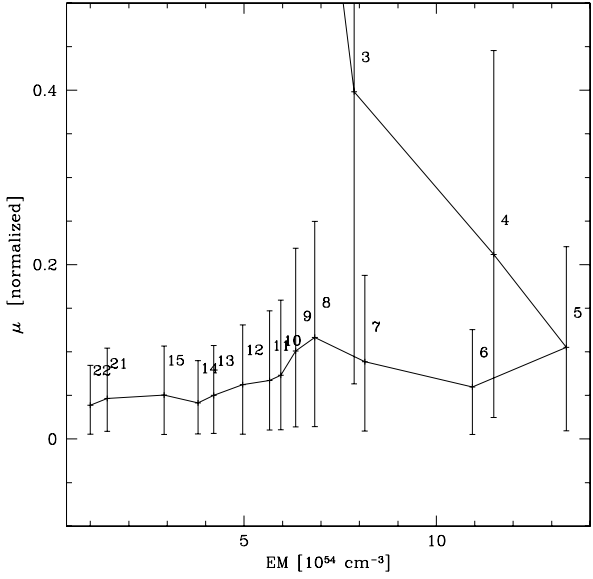, width=8.5cm, bbllx=10pt,
      bblly=150pt, bburx=600pt, bbury=700pt, clip=}
    \caption{The temporal evolution of the $\mu$ parameter (the ratio
      between the radiative and conductive cooling time for the loop
      in quasi-static formalism) during the flare.}
    \label{fig:mu}
  \end{center}
\end{figure}

To determine whether the quasi-static formalism is applicable in
principle, the constancy of the ratio between the radiative and
conductive cooling time during the flare decay has to be checked. This
ratio can be written as (Eq.~(27) of \cite{om89}):
\begin{equation}
  \label{eq:mu}
  \mu = {\tau_{\rm r} \over \tau_{\rm c}} = C \times {T^{13/4} \over
    {EM}} ,
\end{equation}
where $C$ is a constant incorporating all the geometrical factors and
the subscripts to $T$ and $EM$ indicate the power of 10 which the
relevant quantity has been normalized to (in cgs units). In a
quasi-static cooling phase this quantity must be constant.  The time
evolution of $\mu$ for the Algol flare discussed here is plotted in
Fig.~\ref{fig:mu}, using the values of $T$ and $EM$ shown in
Fig.~\ref{fig:tflare}. Once the emission measure has reached its peak
(i.e.\ in time interval~5) $\mu$ is constant within the error bars,
although with some evidence for a slight increase between time
intervals 6 to 8, which may indicate additional heating, as observed
in the time evolution of the plasma temperature (see
Fig.~\ref{fig:tflare}).

\subsubsection{The shape of the decay}

The formalism discussed above implicitly assumes that a decay time can
be uniquely defined, i.e.\ that the relevant quantities decay in a
smooth and monotonic way. For the flare discussed here it is evident
that the decay phase encompasses different phases: a first phase
(intervals 4--5) in which the light curve as well as the derived
temperature and emission measure are decaying exponentially on a
rather fast $e$-folding time scale, a phase in which the decay has
slowed down and in which the temperature is increasing again
(intervals 6--8, obviously indicative of the presence of prolonged
heating) and a final long decay phase in which the light curve decay
is well described by a slow exponential decay, similar to the
temperature and emission measure decays. Thus, the conditions for the
quasi-static formalism are not satisfied for the whole decay phase;
however, it should be applicable to the phase in which the decay is
monotonic and smooth, with an exponential light curve, i.e.\ between
time intervals 8 and 14 inclusive.

\begin{figure}[htbp]
  \begin{center}
    \leavevmode \epsfig{file=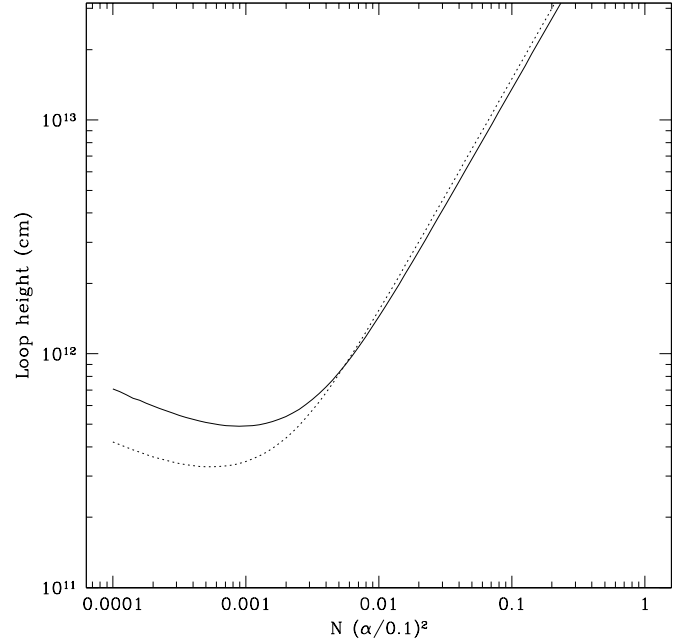, width=9.5cm, bbllx=10pt,
      bblly=150pt, bburx=600pt, bbury=700pt, clip=}
    \caption{The loop's length calculated in the framework of the
      quasi-static cooling formalism, as a function of the quantity
      $N(\alpha/0.1)^2$, where $N$ is the number of flaring loops and
      $\alpha$ is the ratio between the loop's length and its
      diameter. The dashed curve is for loops with a rather strong
      expansion ($\Gamma = 10$), while the continuous curve is for a
      loops without any expansion ($\Gamma = 1$).}
    \label{fig:hn}
  \end{center}
\end{figure}

During this interval the $e$-folding times for the temperature and
count rate of the flare are 97 and 64\,ks respectively, which combine
to yield an effective decay time, as defined above, $\tau_{\rm eff} =
59$\,ks.  The temperature at the beginning of the decay phase thus
defined is 7.2\,keV, or 83\,MK, and the emission measure is $6.4
\times 10^{53}$\,cm$^{-3}$. If we apply the formalism of \cite*{om89},
we can calculate the size of the flaring loop as a function of the
loop expansion factor $\Gamma$, and of the number $N$ and the ratio
$\alpha$ between the loop's length and its diameter at the base. The
resulting relationship between the loop length and $N \alpha^2$ is
plotted in Fig.~\ref{fig:hn}. Unless very small loop diameters are
postulated (i.e.\ smaller than one hundredth of the loop length),
Fig.~\ref{fig:hn} shows that the flaring loop must be longer than a
few times $10^{12}$\,cm (i.e.\ of order 10 stellar radii), independent
from any possible loop expansion.

\begin{figure*}[htbp]
  \begin{center}
    \leavevmode \epsfig{file=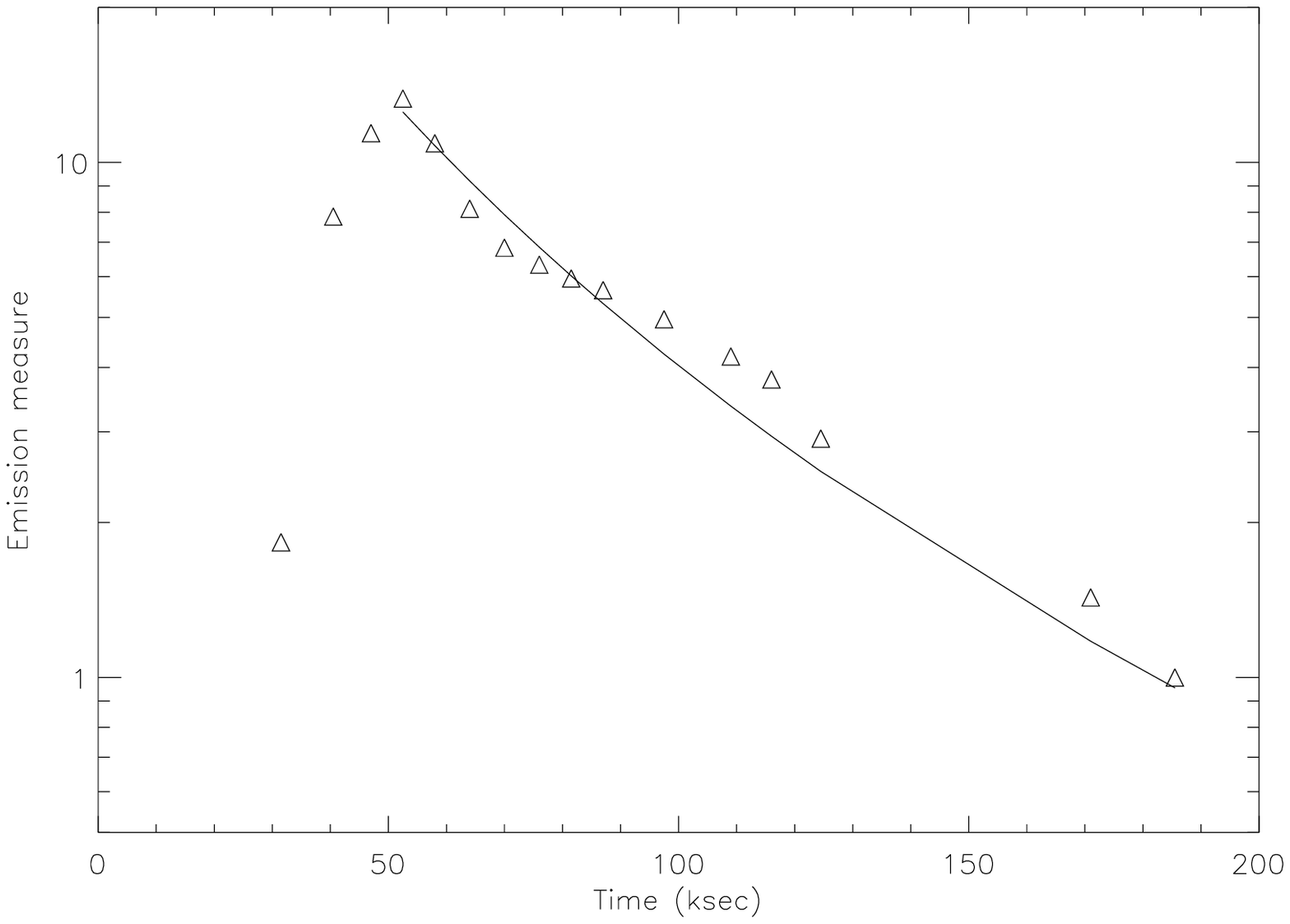, width=8.0cm}
    \leavevmode \epsfig{file=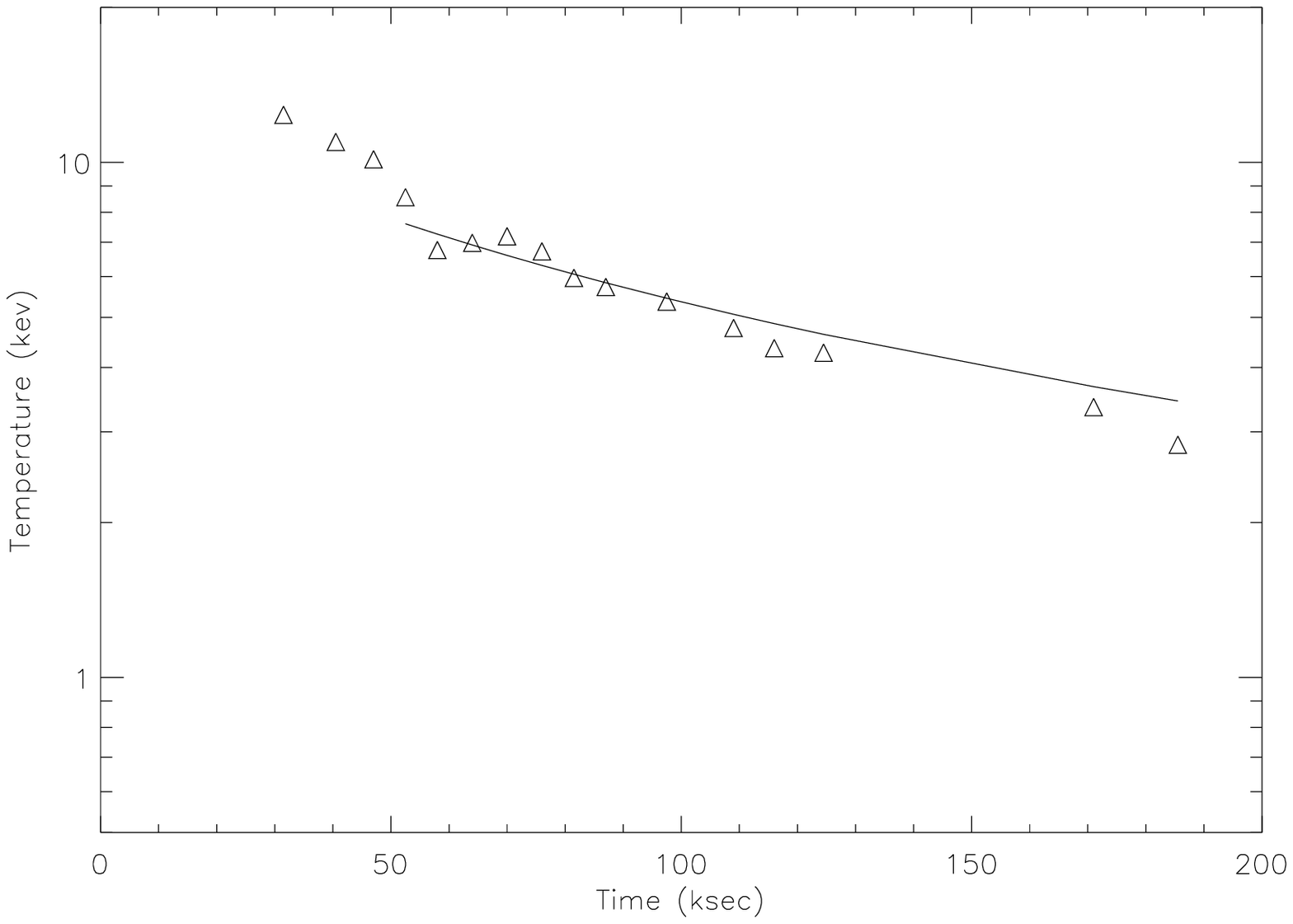, width=8.0cm}
    \caption{The best-fit decay curves for the emission measure (left
      panel) and temperature (right panel) during the flare's decay,
      obtained with a full fit to the quasi-static formalism of van
      den Oord \& Mewe (1989), considering the complete decay of the
      flare. The emission measure is plotted in units of
      $10^{54}$\,cm$^{-3}$}
    \label{fig:emfit}
  \end{center}
\end{figure*}

\subsubsection{Scaling law approximation}

To further constrain the geometry of the flaring loops, we can follow
the approach of \cite*{sut+92}, who show that, once the applicability
of the quasi-static framework has been verified, the geometric
parameters of the flaring loops, as well as the plasma density, can be
derived through simple scaling laws from the detailed analysis of
\cite*{om89} of the EXOSAT flare of Algol (their Eqs.~(19)). The
constancy of the ratio between the radiative and conductive cooling
time gives a scaling law of the form
\begin{equation}
  \label{eq:sca1}
  n_{\rm A} L \propto T_{\rm A}^{(2\gamma +7)/4}
\end{equation}
where the subscript A refers to the apex of the loop, where the bulk
of the emitting plasma is located. Combining the above equation with
the definition of the radiative cooling time $\tau_{\rm r}$
(Eq.~(\ref{eq:taur})), yields $L \propto \tau T^{7/8}$ (again for
$\gamma = -0.25$), and $n \propto \tau^{-1} T^{6/8}$. If we scale from
the analysis of \cite*{om89}, the decay time of the flare observed
here is 10.5 times longer, while the temperature at the beginning of
the decay phase is 1.44 times higher. Thus, the loop length is 14.5
times longer, or $2.3 \times 10^{12}$\,cm, and the density is 8.0
times lower, or $3.3 \times 10^{10}$\,cm$^{-3}$. The remaining
geometrical parameters for the flaring loops can be derived from the
definition of the emission measure
\begin{equation}
  \label{eq:em}
  EM \simeq {\pi \over 8} n_e^2 L^3 (\Gamma +1) N \alpha^2
\end{equation}
where the relationship is approximate because of the (weak) dependence
of the scaling laws on $\Gamma$. Substituting all the values yields
$(\Gamma +1) N (\alpha/0.1)^2 = 1.2 \times 10^{-2}$.

The loop length predicted by the quasi-static formalism for the flare
discussed here is clearly substantial: a length of $2.3 \times 10^{12}$\,cm
corresponds to a loop height of $1.4 \times 10^{12}$\,cm, or $5.7
R_{\rm K}$, large but still within the pressure scale height for the K
star in Algol (for the hot plasma at the beginning of the decay phase
-- 83\,MK -- the pressure scale height for the K star is $\simeq 6
\times 10^{12}$\,cm, or 24\,$R_{\rm K}$).

If the decay time constants are derived from the complete decay phase,
i.e.\ from the moment in which both $T$ and $EM$ start to decrease
(interval 5) the results do not change significantly. In this case the
precise form of the decay is ignored, and only the time scales are
considered. This approach is also likely to be more consistent with
the majority of the published results on stellar flares, in which
neither the temporal coverage nor the statistics are sufficient to
show the details of the time evolution of the flare parameters, and a
single determination of the decay time constants starting from the
flare peak is done. For example, for the Algol flare observed by
ROSAT, only a sporadic coverage of the flare light curve was
available, so that any short-lasting intermediate heating episodes
such as the ones evident in this case would not have been detected.

In the present case the $e$-folding times (derived from the peak of
the quantity of interest) for the temperature and count rate are,
respectively, 118 and 59~ks, combining to yield an effective decay
time of 63~ks, and the temperature at the beginning of the decay is
98~MK. Application of the same scaling laws as above results in a loop
length of $2.8 \times 10^{12}$\,cm and a plasma density of $3.5 \times
10^{10}$\,cm$^{-3}$, essentially identical to the results obtained by
considering only the exponential part of the decay.

\subsubsection{Full fits to the scaling-law formalism}

We have also performed a full fit of the equations of \cite*{om89} to
the complete decay phases of the temperature and count rate. The
resulting best-fit is shown in Fig.~\ref{fig:emfit}. The resulting
loop length is $1.7 \times 10^{12}$\,cm with a corresponding plasma
density of $\simeq 10^{10}$\,cm$^{-3}$. Again, the resulting loop
length and density are quite similar to the value obtained with a
simple application of the scaling laws to the exponentially decaying
part of the light curve, confirming the relative insensitivity of the
quasi-static formalism to the details of the starting assumptions.

\subsection{The slope in the temperature-density diagram}
\label{sec:zeta}

A different approach to the analysis of the decay times of spatially
unresolved flares has been developed by \cite*{rbp+97}.  This method
simultaneously yields estimates for the physical size of the flaring
loop as well as for the presence and time scale of heating during the
decay phase of the flare. The method uses the slope of the locus of
points in the temperature versus density diagram during the flare
decay phase (\cite{sss+93}) as a diagnostic of the presence of
sustained heating.  Under the assumption that the loop volume remains
constant during the flare, the square root of the emission measure is
used as a proxy for the density; we will refer to this approach, in
the following, as the $EM$ vs. $T$ slope method.

Detailed hydrodynamical simulations show that flares decay
approximately along a straight line in the $\log \sqrt{EM}$ -- $\log
T$ diagram, and that the value of the slope $\zeta$ of the decay path
is related to the ratio between the observed decay time of the light
curve $\tau_{\rm lc}$ and the ``natural'' thermodynamic cooling time
of the loop without additional heating $\tau_{\rm th}$.  The validity
of the simulation-derived relationship has then been verified by
\cite*{rbp+97} with a sample of solar flares observed by Yohkoh. This
approach allows to explicitly estimate the intrinsic spread of the
derived physical parameters; in practice the estimated loop size
agrees, for the solar case, to $\simeq 20\%$ with the actual size. The
recalibration of the method for temperatures and emission measures
derived with the BeppoSAX MECS detector is discussed by (Pallavicini
et~al. in preparation).

In practice, an observed flare decay with a slope $\zeta \simeq 1.7$
(again for $T$ and $EM$ derived with the MECS detector) implies that
no additional heating is present ($\tau_{\rm h} = 0$) while smaller
values of $\zeta$ imply progressively longer heating time scales.  The
relationship between $\zeta$ and $\tau_{\rm lc}/\tau_{\rm th}$ becomes
effectively degenerate at values of $\zeta$ smaller than about 0.4, so
that $\zeta \la 0.4$ implies a heating time scale comparable to the
observed decay time of the flare (see Fig.~2 of \cite{rbp+97}). The
loop size is estimated as a function of $\tau_{\rm lc}$, $T_{\rm max}$
and $\zeta$ (where $T_{\rm max}$ is the peak flare temperature, not
the temperature at the beginning of the decay phase).  For a given
$\tau_{\rm lc}$ and $T_{\rm max}$, the smaller $\zeta$ (and thus the
longer the additional heating) the smaller the implied loop length.
In a large fraction of the solar flares examined by \cite*{rbp+97}
significant heating is present, so that the thermodynamic decay time
of the loop alone significantly over-estimates its size (by factors
between 2 and 10).

\begin{figure}[htbp]
  \begin{center}
    \leavevmode \epsfig{file=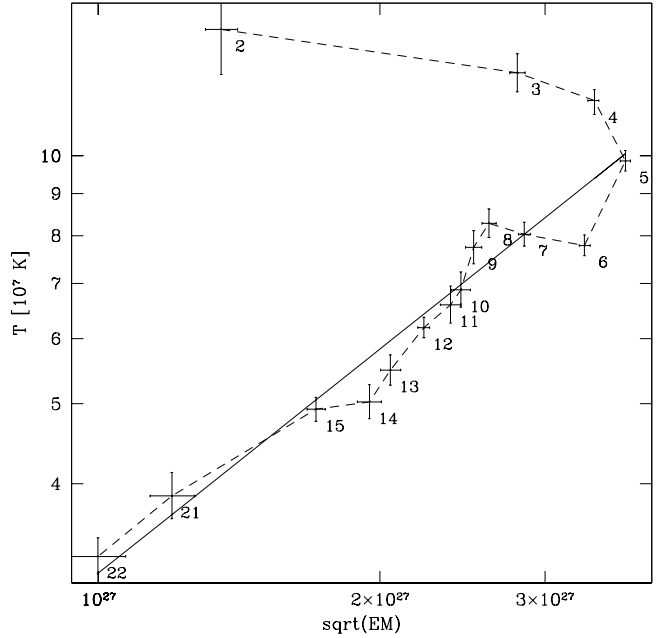, width=9.5cm, bbllx=5pt,
      bblly=150pt, bburx=600pt, bbury=700pt, clip=}
    \caption{The evolution of the flare temperature (in K) and
      square root of the emission measure (in cm$^{-3}$, a proxy to
      the plasma density, in the assumption of a constant volume). The
      error bars are $1\,\sigma$ for three interesting parameters
      ($\Delta \chi = 3.50$), and the dashed line connects each
      individual point, to guide the eye.  Also indicated is the
      correspondence with the time intervals defined in
      Fig.~\ref{fig:lc}. The continuous line is the fit to the points
      from 8 to 22.}
    \label{fig:zeta}
  \end{center}
\end{figure}

The evolution of the spectral parameters for the Algol flare is
plotted, in the $\log \sqrt{EM}$ -- $\log T$ plane, in
Fig.~\ref{fig:zeta}. The set of $\sqrt{EM}$, $\log T$ pairs plotted
includes the rising phase of the flare (up to interval 5), when the
emission measure is still building up, and thus the decay has not yet
begun. Afterwards the initial very steep decay only lasts for a
relatively brief time, as the temperature increases again in intervals
6 to 8. The apparently undisturbed decay begins, as discussed in the
framework of the quasi-static method, with interval 8, and, up to
interval 14, follows a clean straight line. The last three intervals
(which however do not sample very well the light curve, being
interrupted by the eclipse) show evidence for a change in the slope,
which becomes shallower.

\cite*{rbp+97} quote four conditions for the method to yield optimal
results, i.e.\ the slope $\zeta$ must be greater than $0.3$, the light
curve must decay exponentially, the trajectory of the decay in the
$\log T$ -- $\log \sqrt{EM}$ plane must be linear and the resulting
loop length must well below the local pressure scale height.  In the
present case the decay is indeed not linear, but there is clear
evidence for an episode of reheating, so that the slope $\zeta$ is not
very well constrained, and the resulting uncertainties are likely to
be significantly larger.  The choice of which interval of the observed
decay to use for the determination of the flare parameters is, in this
case, prescribed by the method; it begins when the emission measure
peaks and ends with the time interval in which the light curve has
decayed to a count rate of 10\% of the peak value.

The average slope of the decay in the $\log T$ -- $\log \sqrt{EM}$
plane is $0.84 \pm 0.20$ (90\% uncertainty, as in all of the
following). By applying the equivalent of Eq.~(3) of \cite*{rbp+97},
the ratio between the observed light-curve decay time and the
thermodynamic decay time of the flaring loop (i.e.\ the decay time
which would be observed in the absence of heating) is given by
(recalibrated for the BeppoSAX MECS detector)
\begin{equation}
  \label{eq:zeta}
  \tau_{\rm LC}/\tau_{\rm th} = 8.68 \times e^{-\zeta/0.59} + 0.3 =
  F(\zeta)
\end{equation}

In our case, $\tau_{\rm LC}/\tau_{\rm th} = 2.4$. The loop length is
given by the equivalent of Eq.~(4) of \cite*{rbp+97}, i.e.\ 
\begin{equation}
  \label{eq:lzeta}
  L = {{\tau_{\rm th}\sqrt{0.233 \times T_{\rm p}^{1.099}}} \over
    {3.7\times 10^{-4}}}
\end{equation}
where $T_{\rm p}$ is the measured peak flare temperature. Substituting
the values determined for the Algol flare, i.e.\ $\tau_{\rm LC} = 49.6
\pm 4.5$~ks (determined from the peak down to the 10\% level),
$\tau_{\rm th} = 20.1$~ks, $T_{\rm p} = 138$~MK, the predicted loop
length is $L = 8.2\times 10^{11}$~cm ($3.3~R_{\rm K}$), with a 90\%
uncertainty on the derived length is $\pm 3.5 \times 10^{11}$~cm,
yielding a formally allowed range of loop lengths of 4.7 to $11.7
\times 10^{11}$~cm, i.e.  1.9 to $4.7~R_{\rm K}$. The peak temperature
at the loop apex is $T_{\rm max} = 200$~MK. As expected for a flare
with significant heating, the loop length derived is significantly
shorter (a factor of 2 to 4) than the value derived through the
quasi-static formalism under the assumption of no heating.

No density estimate is produced directly by the \tems\ analysis. A
simple-minded density estimate can be derived by computing the loop's
volume (with a given assumed value of $\alpha$, typically $\simeq
0.1$) and computing the peak density from the peak emission measure
through the simple relationship
\begin{equation}
  \label{eq:dens}
  n \simeq \sqrt{EM \over V} \simeq \sqrt{EM \over 2 \pi \alpha^2 L^3}
\end{equation}

The density corresponding to the range of loop lengths derived through
the \tems\ method is $2.7 \times 10^{9} / \alpha$ to $1.9 \times
10^{9} / \alpha$~cm$^{-3}$.

\section{Comparison with the flaring region size determined by
  eclipse analysis}
\label{sec:length}

The lengths derived through the analysis of the flare decay range
between $4.7\times 10^{11}$~cm (the lower end of the range of values
allowed by the \tems\ method) and $28 \times 10^{11}$~cm (the longer
value obtained with the quasi-static formalism); the corresponding
loop heights (under the assumption of a simple, single-loop geometry)
above the stellar surface range from $\simeq 1.2$ to $\simeq 7$ radii
of the K star. The corresponding plasma densities are few times
$\simeq 10^{10}$~cm$^{-3}$.

How realistic are the long tenuous loops implied by the quasi-static
decay-phase analysis? Unlike any previous stellar flare, we can in
this case compare these values with the allowed volume for the flaring
plasma derived from a geometrical analysis of the eclipse of the X-ray
emission. As discussed in detail in \cite*{sf99}, the observed total
eclipse of the flaring plasma implies a maximum height of the flaring
region of less than $\simeq 0.6$ stellar radii, i.e. $h \le 0.6~R_{\rm
  K} \simeq 1.5 \times 10^{11}$~cm, equivalent to a maximum loop
length (assuming a vertical loop with a simple geometry) of $\le 2.4
\times 10^{11}$~cm, significantly smaller than the values derived by
the analysis of the quasi-static flare decay. The lower end of the
allowed range for the loop lengths derived with the \tems\ method is
closer to the geometrical size derived through the eclipse analysis,
but still marginally too large.

An immediate consequence of the small volume implied by the eclipse is
that the flaring plasma must have a higher density than predicted by
the quasi-static analysis. The density is estimated by \cite*{sf99} at
$f^{-0.5} \times 10^{11}$~cm$^{-3}$, where $f$ is the filling factor
for the flaring plasma within the region allowed by the eclipse, $f <
1$. The high density in turn implies a decay time for radiative
cooling for the plasma of $24 f^{0.5}$~ks, and a conductive decay time
only $2.5 f^{-0.5}$~ks, with a resulting effective decay time, for any
value of $f$, negligible when compared with the observed long decay
time of the flare.  Therefore, the observed decay must be driven
almost entirely by the evolution of the energy dissipation responsible
for the heating. In this framework, the observed irregularity of the
observed flare decay is then not so much related to details of the
decay process or of the flaring structure, but rather to the temporal
evolution of the energy dissipation process.

The large loop sizes derived from the quasi-static analysis of flare
decay, although somewhat extreme, are not unusual: large loop lengths
are a common result from the decay-phase analysis of large flares
observed on all classes of coronal sources.  For example,
\cite*{sch94} shows that, using the quasi-static cooling paradigm, for
the flare on the dMe star EV~Lac observed during the ROSAT All Sky
Survey the implied loop length is $6 \times 10^{11}$~cm (i.e.\ $\simeq
10$ stellar radii); recently, \cite*{tkm+98} reported the detection of
a large flare on the pre-main sequence star V773~Tau, showing that the
observed cooling time implies loop sizes of $\simeq 4 \times
10^{11}$~cm, or $\simeq 1.2$ stellar radii; the very long ROSAT flare
observed on the active binary CF~Tuc (\cite{ks96}) implies (depending
on the details of the assumptions) loop lengths of order $2$ to $5
\times 10^{11}$~cm, again comparable to or larger than the stellar
radius.  Such large loop sizes are in contrast with the solar case,
where, even for the strongest flares, the size of the flaring loops
remains well below the solar radius. While the energy released for the
stellar flares discussed here is several orders of magnitude larger
than in the case of the strongest solar flares, and thus the solar
analogy may not necessarily be fully relevant, the very large loops
raise several difficulties related to the high implied magnetic fields
at large distances from the stars, the heating mechanisms, etc.  While
the presence of long magnetic structures anchored on both stars could
somewhat alleviate these concerns in the case of active binary systems
(but not for Algol-type systems), this would not work in the case of
single stars.

Thus the physical meaning of the long loop sizes derived for
spatially unresolved stellar flares from the quasi-static analysis of
the loop decay appears questionable when compared with the geometric
constraints on the size of the flaring region derived from the eclipse
analysis. The present observations show that large flares can be
produced by quite compact, high density structures, much smaller than
implied by the quasi-static analysis of the loop decay phase or by
simple considerations about the radiative decay time of an undisturbed
plasma.  Sustained heating, fully dominating the observed decay, must
therefore, in such cases, be present. The heating mechanism must then
not be impulsive only, but must be followed by a long-lasting tail,
with time scales of order of tens of ks. The \tems\ analysis in this
case does not yield a very stringent limit to the size of the flaring
loop { (and it still appears to marginally overestimate the loop's
  size)}, but provides clear support for the presence of long-lasting
heating. 

The applications of the \tems\ analysis in a situation in which the
flare decay in the $\log T$ -- $\log \sqrt{EM}$ plane is not linear is
likely to result in larger uncertainties on the derived parameters.
The flare shows clear evidence for a significant reheating episode (in
time intervals 6 to 8), which alters the decay path, unlikely the
hydrodynamic simulations which are used to derive the loop's size, in
which the heating is parameterized as a monotonically decreasing
function of time. Note that if the same flare had been observed at a
significantly lower $S/N$, such reheating episode would not have been
resolved, but it would simply have resulted in a linear but shallower
flare decay, biasing the resulting flare parameters. One alternative
possibility is that the flare is the result of a different physical
process that the single, constant-volume loop (a ``compact'' flare)
which is assumed throughout the analysis. This could be, e.g. --
similarly to the solar ``two-ribbon'' flares -- an evolving loop
arcade. Once more this would imply that different physical conditions
compared to the hydrodynamic simulations of \cite*{rbp+97}, and thus
larger uncertainties on the results.

\subsection{Comparison with previous flares observed on Algol}

A few major flares have already been observed on Algol by a variety of
X-ray telescopes. The physical parameters derived from the analysis of
the decay phase are listed in Table~\ref{tab:comp}.

\begin{table}[htbp]
  \begin{center}
    \caption{A comparison of the physical parameters derived for
      different flares on Algol from the GINGA, EXOSAT and ROSAT
      flares. The first column gives the maximum temperature measured
      in the flare, the second column the temperature at the beginning
      of the decay phase, the third column the maximum emission
      measure measured during the flare.  The fourth column gives the
      effective decay time for the flare, while the remaining two
      columns give the plasma density and the loop length derived from
      the quasi-static analysis. Numerical subscripts indicate the
      power of 10 by which the relevant quantity has been scaled.
      }
    \leavevmode
    \begin{tabular}{lrrrrrr}
      Instr. & $T_{\rm p}$ & $T_{\rm d}$ &$EM_{54}$ & $\tau_{\rm eff}$
      & $n_{9}$ & $L_{10}$ \\
                 & MK & MK & \,cm$^{-3}$ & ks &
      cm$^{-3}$ & cm \\\hline
      GINGA      &  67 & 65 & 1.34 & 22.5 &  50 &  60 \\
      EXOSAT     &  80 & 58 & 0.98 &  5.2 & 260 &  16 \\
      ROSAT      &  88 & 39 & 10.0 & 15.4 & 170 &  12 \\
    \end{tabular}
    \label{tab:comp}
  \end{center}
\end{table}

\begin{table}[htbp]
  \begin{center}
    \caption{A comparison of the physical parameters derived for
      the BeppoSAX flare of Algol, derived using different methods and
      assumptions. ``QS-SL-EX'' indicates application of the
      quasi-static formalism to the exponentially decaying part of the
      light curve, and using the scaling laws of Stern et~al.\ (1992)
      to obtain the loop parameters; ``QS-SL-FD'' indicates the same
      approach applied to the full decay phase; ``QS-FF-FD'' indicates
      application of the quasi-static approach, with a full-fit of the
      complete decay phase to the equations of van den Oord \& Mewe
      (1989); ``R+97-FD'' indicates application of the Reale et~al.\ 
      (1997) formalism to the complete decay phase, while the row
      marked with ``Eclipse'' indicates the parameters derived from
      the analysis of the eclipse light curve by Schmitt \& Favata
      (1999). In this last case only the loop length and density are
      derived; the density includes the unknown filling factor $f$.
      Numerical subscripts indicate the power of 10 by which the
      relevant quantity has been scaled.}
    \leavevmode
    \begin{tabular}{lrrrrrr}
      Method & $T_{\rm p}$ & $T_{\rm d}$ &$EM_{54}$ & $\tau_{\rm eff}$
      & $n_{9}$ & $L_{10}$ \\
                 & MK & MK & \,cm$^{-3}$ & ks &
      cm$^{-3}$ & cm \\\hline
      QS-SL-EX   & 140 & 83 & 13.3 & 59.0 &  33 & 230 \\
      QS-SL-FD   & 140 & 98 & 13.3 & 63.0 &  35 & 280 \\
      QS-FF-FD   & 140 & 90 & 13.4 & 43.5 &  90 & 174 \\
      R+97-FD    & 140 & 140 & 13.3 & 22.6 & $\simeq 2/\alpha$ &  47--120 \\
      Eclipse & 140 & -- & -- & 13.3 & $94/\sqrt{f}$ &  24 \\
    \end{tabular}
    \label{tab:compsax}
  \end{center}
\end{table}

For comparison we also list, in Table~\ref{tab:compsax}, the parameter
values derived, for the BeppoSAX flare, with the different
approaches discussed above.

\section{Conclusions}
\label{sec:concl}

The large Algol flare discussed here has several unique
characteristics, already listed in the Introduction, which make it a
unique event given the level of detailed constraints which can be
derived on the flaring process. The most important conclusions are:

\begin{itemize}

\item The variations of metallicity of the flaring plasma (which had
  already been hinted at in the study of previous X-ray flares) can be
  unambigously determined. The long duration of the flare and the good
  count statistics make it possible to determine the temporal
  evolution of the metal abundance, showing that it rises by a factor
  of three during the flare rise and decays again to the pre-flare
  value on time scales faster than any of the observed time scales
  (for the light curve, emission measure of temperature decay).
  Alternative explanations (as opposed to real abundance variation
  effects) for the observed changes in the spectrum, such as
  non-equilibrium effects, are difficult to reconcile with the plasma
  density for the flaring loop (a fortiori for the higher densities
  implied by the eclipse), which imply that the plasma would relax to
  equilibrium ionization conditions in a few tens of seconds at most.
  Fluorescence from the X-ray bombarded photosphere is in contrast
  with the lack of observed shift of the centroid of the Fe~K line
  toward 6.4~keV and from the fact that the flare occurs in the
  permanently-occulted pole of the K-type star, so that the
  potentially fluorescing photosphere would in any case be largely
  self-eclipsed.

\item During the initial (rising) phase of the flare the best-fit
  absorbing column density is large ($\simeq 3 \times
  10^{21}$~cm$^{-2}$), and decays on time scales of some tens of ks.
  We interpret this as possibly associated with moving, cool absorbing
  material in the line of sight, i.e.\ a major coronal mass ejection
  associated with the flare's onset.

\item The length derived for the flaring loop from the analysis of the
  flare decay, with the quasi-static formalism of \cite*{om89} is
  consistently too large when compared with the upper limit on the
  size of the flaring region imposed by the observed total eclipse
  (see \cite{sf99}).  Different assumptions result in a small range of
  derived loop half-lengths (from $18$ to $28 \times 10^{11}$~cm),
  they are all well above the eclipse-derived upper limit of $2.4
  \times 10^{11}$~cm. Given the nature of the geometric constraints
  imposed by the eclipse observations, the conclusions is thus that
  the decay-derived loop lengths are too large, by a factor of at
  least a few times, and the derived densities are correspondingly
  lower. This may be relevant also for the interpretation of the very
  large loop sizes which have been determined with this method in the
  past for other large X-ray flares on other stars.
  
\item The range of loop lengths derived with the analysis method of
  \cite*{rbp+97} imply the presence of significant sustained heating,
  in agreement with the conclusions drawn on the basis of the
  eclipse-derived size.  The derived lengths are still marginally too
  large with respect to the eclipse-derived size; this is likely due
  to the application of the method in the presence of a significant
  reheating episode, which (when compared with hydrodynamical
  simulations with monotonically decreasing heating) will introduce
  additional uncertainties in the results.  Nevertheless, even in such
  case, this method appears to yield more reliable information on the
  physical conditions of the flaring region than the quasi-static
  method.

\item As a consequence of the smaller loop length implied by the
  eclipse, sustained heating must be present throughout the flare, and
  it must actually be driving the decay. Given the small intrinsic
  thermodynamic decay time of the loop implied by the small size and
  corresponding high density, the loop has a small ``thermal
  inertia'', and thus the heating time profile must essentially be the
  same as the observed X-ray luminosity profile (given in
  Fig.~\ref{fig:energy}).

\end{itemize}

\begin{acknowledgements}

  We would like to thank M.~Guainazzi and D.~Dal~Fiume for their help
  in the analysis and understanding of the PDS data, and A.~Parmar for
  the useful discussion on the LECS data. We are grateful to F.~Reale
  for the several illuminating discussions on his methodology for the
  analysis of flare decays as well as for calibrating his method for
  the BeppoSAX MECS detectors. The BeppoSAX satellite is a joint
  Italian and Dutch program.

\end{acknowledgements}


\end{document}